\documentclass[useAMS,usenatbib]{mn2e}

\usepackage{graphicx}	
\usepackage{amsmath}	
\usepackage{amssymb}	

\title[A 10\ kpc stellar substructure in the outer LMC]{A 10\ kpc stellar substructure at the edge of the Large Magellanic Cloud: perturbed outer disk or evidence for tidal stripping?}

\author[A. D. Mackey et al.]{A. D. Mackey$^{1}$\thanks{E-mail: dougal.mackey@anu.edu.au},  S. E. Koposov$^{2}$, D. Erkal$^{2}$, V. Belokurov$^{2}$, G. S. Da Costa$^{1}$
\newauthor and F. A. G\'{o}mez$^{3}$
\\
$^1$Research School of Astronomy and Astrophysics, Australian National University, Canberra, ACT 2611, Australia\\
$^2$Institute for Astronomy, University of Cambridge, Madingley Road, Cambridge, CB3 0HA, UK\\
$^3$Max-Planck-Institut fuer Astrophysik, Karl-Schwarzschild-Str. 1, D-85748, Garching, Germany
}

\date{Draft version \today.}

\pubyear{2016}

\begin{document}
\label{firstpage}
\pagerange{\pageref{firstpage}--\pageref{lastpage}}
\maketitle

\begin{abstract}
We report the discovery of a substantial stellar overdensity in the periphery of the Large Magellanic Cloud (LMC), found using public 
imaging from the first year of the Dark Energy Survey.  The structure appears to emanate from the edge of the outer LMC disk at a 
radius $\approx 13.5\degr$ due north of its centre, and stretches more than $10$\ kpc towards the east. It is roughly $1.5$\ kpc 
wide and has an integrated $V$-band luminosity of at least $M_V = -7.4$. The stellar populations in the feature are indistinguishable 
from those in the outer LMC disk. We attempt to quantify the geometry of the outer disk using simple planar models, and find that 
only a disk with mild intrinsic ellipticity can simultaneously explain the observed stellar density on the sky and the 
azimuthal line-of-sight distance profile. We also see possible non-planar behaviour in the outer disk that may reflect a warp and/or
flare, as well as deviations that resemble a ring-like structure between $\sim9\degr-12\degr$ from the LMC centre. Based on all these 
observations, we conclude that our remote, stream-like feature is likely comprised of material that has been stripped from the
outskirts of the LMC disk, although we cannot rule out that it represents a transient overdensity in the disk itself. We conduct a simple 
$N$-body simulation to show that either type of structure could plausibly arise due to the tidal force of the Milky Way; however we also
recognize that a recent close interaction between the LMC and the SMC may be the source of the stripping or perturbation.
Finally, we observe evidence for extremely diffuse LMC populations extending to radii of $\sim 18.5$\ kpc in the disk plane 
($\approx 20\degr$ on the sky), corroborating previous spectroscopic detections at comparable distances.
\end{abstract}




\section{Introduction}
The Large and Small Magellanic Clouds (LMC and SMC) are the nearest examples of star-forming, gas-rich dwarf galaxies.  
They sit well within the halo of the Milky Way, at Galactocentric distances of $\approx 50$\ kpc and $\approx 60$\ kpc, respectively.  
Their proximity renders them ideal targets for investigating a huge variety of different astrophysical problems; their stellar populations
can be easily resolved with ground-based facilities, and studied in exquisite detail. 

A critical ingredient in understanding many of the peculiar features of the Magellanic System -- for example the striking off-centre bar 
in the LMC \citep[e.g.][]{devau:72,vdm:01b}, the broad age gap in the LMC star cluster system \citep[e.g.,][]{dacosta:91}, the vigorous ongoing 
star formation in both galaxies, the irregular morphology of the SMC \citep[e.g.,][]{westerlund:97}, and the Magellanic Stream and Bridge of 
H{\sc i} gas \citep[e.g.,][]{mathewson:74,putman:98,putman:03,nidever:08,nidever:10}-- is the interaction history of the LMC and SMC, both
with each other and with the Milky Way.  This is still not well constrained.  It has only been during the last decade, with precision proper
motion measurments from the {\it Hubble Space Telescope} that evidence has mounted that the Clouds have only recently fallen deep
into the gravitational potential of the Milky Way \citep[e.g.,][]{kalli:06,kalli:13}, and may indeed be on their first close passage. As a result 
it is possible that close encounters between the LMC and SMC have been more influential in shaping their evolution and appearance 
\citep[e.g.,][]{besla:12,diaz:12}. 

The outskirts of the LMC and SMC are of particular interest. These low-density regions are most susceptible to disturbances due to 
interactions between the two galaxies and due to the tidal force of the Milky Way.  Key questions are the extent to which the structure
of the outer LMC disk exhibits evidence of distortion, the nature and extent of any stellar component stretching between the Clouds in the 
Magellanic Bridge region, whether there is a stellar component associated with the Magellanic Stream, and whether either of the Clouds 
possesses a spheroidal halo component or shows evidence for the presence of disrupted satellites. The latter may be expected because 
in the $\Lambda$CDM cosmological model structures are formed hierarchically at all scales, such that a number 
of the dwarfs in the Local Group could have experienced significant merger events within the past several Gyr \citep[e.g.,][]{deason:14}. 

Studying the remote peripheries of the LMC and SMC is observationally extremely challenging, requiring exceptionally low surface brightness
features ($V \ga 30$ mag arcsec$^{-2}$) to be traced over large swathes of sky. Although early studies utilising photographic plates were
able to explore the morphology of the Clouds at intermediate radii \citep[e.g.,][]{devau:72,irwin:91}, it is only quite recently, with the advent
of wide-field imagers and multi-object spectrographs mounted on medium and large aperture telescopes in the southern hemisphere, that 
the problem of tracing the LMC and SMC to their outer limits has become tractable. 

The structure of the LMC is generally accepted to be that of a planar disk viewed at an angle inclined to the plane of the sky.
Pioneering work on the shape, structure and extent of the LMC disk was conducted by \citet{vdm:01a} and \citet{vdm:01b} using 
newly-available {\it contiguous} near-infrared data sets from 2MASS \citep{skrutskie:06} and DENIS \citep{epchtein:97}, that provided, for 
the first time, the opportunity to map the outer regions of the LMC via resolved red giant branch (RGB) and asymptotic giant branch (AGB) 
star counts -- i.e., using (old) stellar populations characteristic of the LMC periphery. These studies showed that the outer disk, or at least 
its old component, appears rather smooth all the way to the edge of the region studied at $\sim 7\degr$ from the LMC centre, and found a 
best-fit inclination angle of $i = 34.7\degr \pm 6.2\degr$ and exponential scale 
length of $\approx 1.3-1.5$ kpc. These values are consistent with those obtained by other studies conducted before and since at more central 
radii using similar tracers \citep[e.g.,][]{weinberg:01,olsen:02,rubele:12} as well as a variety of alternatives (such as Cepheids and H{\sc i} gas), 
and incorporating both photometric and kinematic techniques \citep[see the extensive discussion in][]{vdm:01a}. In addition, \citet{vdm:01a} 
observed tentative
evidence for a decrease in the inclination angle with galactocentric radius, together with a twisting of the position angle of the line of
nodes -- both possibly indicative of warping of the LMC disk at large radii. Moreover, \citet{vdm:01b} found that the LMC disk likely
has an intrinsic elongation, perhaps induced by the tidal force of the Milky Way.

More recently \citet{besla:16} have presented a panoramic view of the LMC in diffuse (unresolved) light out to $\sim 8\degr$ using long 
exposures from small telescopes. They found evidence for faint stellar arcs and multiple spiral arms in the northern part of 
the disk, but with no comparable counterparts in the south. Their numerical models suggest this type of asymmetric arrangement can arise
as the result of repeated close interactions with the SMC.

The extent and structure of the LMC beyond the edge of the region studied by \citet{vdm:01a}, \citet{vdm:01b}, and \citet{besla:16} at 
$\approx 7-8\degr$ 
is not well explored or understood. On the one hand, spectroscopic measurements covering a significant number of disjoint fields at a 
wide variety of azimuthal angles and radial distances report kinematic detection of LMC RGB stars out to $\sim 20\degr$
\citep{majewski:99,majewski:09}; moreover, LMC populations have also been clearly detected, both kinematically and photometrically, 
in the foreground of the Carina dwarf at a radial distance of $\sim 22\degr$ from the LMC centre \citep{majewski:00,munoz:06,mcmonigal:14}.
\citet{majewski:09} hypothesise that these remote stars are indicative of a classical pressure-supported stellar halo surrounding the LMC.
On the other hand, deep photometric studies tracing old main sequence turn-off stars in stripes to the north \citep{saha:10} and 
north-east \citep{balbinot:15} observe a clear exponential decrease in surface density out to $\sim 15\degr-16\degr$, suggesting that 
the smooth LMC disk extends at least 10 scale-lengths from its centre. Beyond this radius, neither work observes definitive evidence for LMC
populations, and each posits that the LMC must be truncated, perhaps tidally, at approximately this point. Most recently, \citet{belokurov:16}
have probed to even larger radii in the north and east using blue horizontal branch stars, and report the discovery of an apparently rather 
substructured distribution reminiscent of a stellar halo extending to at least $\sim 30\degr$.

In this paper we present a detailed investigation of the northern outskirts of the LMC, utilising images taken during the first year of 
the Dark Energy Survey (DES) and released publically through the NOAO science archive. In particular, we explore the geometry and extent 
of the LMC disk and periphery, and report on the discovery of a substantial stream-like stellar substructure situated to the north and 
north-east of the LMC, and stretching to a radius of at least $20\degr$ from the galaxy's centre.

\section{Data}
\label{s:data}
Our analysis is based on photometry derived from publically available images obtained in the first year of DES operations.
DES \citep{abbott:05} utilises the $520$ megapixel Dark Energy 
Camera \citep[DECam;][]{flaugher:15} mounted on the 4m Blanco Telescope at the Cerro Tololo Inter-American Observatory in Chile.  
The DECam imager consists of $62$ CCDs arranged into a roughly hexagonal mosaic spanning a total field of view of 
$\approx 3$\ deg$^2$ with a pixel scale of $0.26\arcsec$\ pixel$^{-1}$.
The DES survey area will ultimately spread over $\sim5000$\ deg$^2$ of the southern sky in {\it grizY}; the first-year 
imaging spans around $\sim 2200$\ deg$^2$ with mostly single-epoch coverage.  This includes a region sitting 
predominantly to the north and west of the LMC, covering position angles between $\sim-90\degr$ and $\sim 45\degr$ 
at radii typically outside $7.5\degr$ but reaching as close as $\approx 6\degr$ to the LMC centre. Overall, roughly 
three-eighths of the LMC disk beyond $7.5\degr$ is covered by the year one DES imaging.

We utilise the photometric catalogue described by \citet{koposov:15}, and refer the interested reader to that paper for a
full description of its construction. In brief, source detection and measurement across all images was conducted using the 
{\it SExtractor} and {\it PSFEx} software \citep[e.g.,][]{bertin:96,bertin:11}.  The individual detection lists were calibrated
to the SDSS scale using DR7 of the APASS survey, cross-matched to remove duplications, and merged across different 
pass-bands into a final catalogue. Star-galaxy separation was undertaken using the {\sc spread\_model} and 
{\sc spreaderr\_model} parameters provided by {\it SExtractor} \citep[as defined in][]{desai:12}. General improvements were made 
to the processing over that reported by \citet{koposov:15}, including an enhanced photometric calibration utilising an 
``uber-calibration'' procedure \citep{pad:08}, extension of the processing to incorporate $z$-band data, and the use of all 
DES imaging from the 2013-2014 season (DES Y1) resulting in somewhat increased areal coverage.

For the present analysis, only objects with clean stellar detections in the $g$, $r$, and $i$ filters were retained.
This results in slightly brighter completeness limits than those reported by \citet{koposov:15} -- by examining the 
position of the peak in source number counts we estimate that the catalogue starts to become significantly affected 
by incompleteness around $g\sim23.0$, $r\sim22.9$, and $i\sim22.2$. 

For all sources, foreground reddening was corrected using the maps of \citet*{schlegel:98} and the extinction coefficients
from \citet{schlafly:11} appropriate for the SDSS photometric system -- i.e., $3.303$, $2.285$, and $1.698$ for the 
$g$, $r$, and $i$ bands respectively. In the direction of the northern outskirts of the LMC, the reddening is very mild with
a median colour excess of $E(B-V) \approx 0.05$.

\section{Analysis}
\subsection{Color-magnitude diagrams}
\label{ss:cmds}
\begin{figure*}
\begin{center}
\includegraphics[height=82mm]{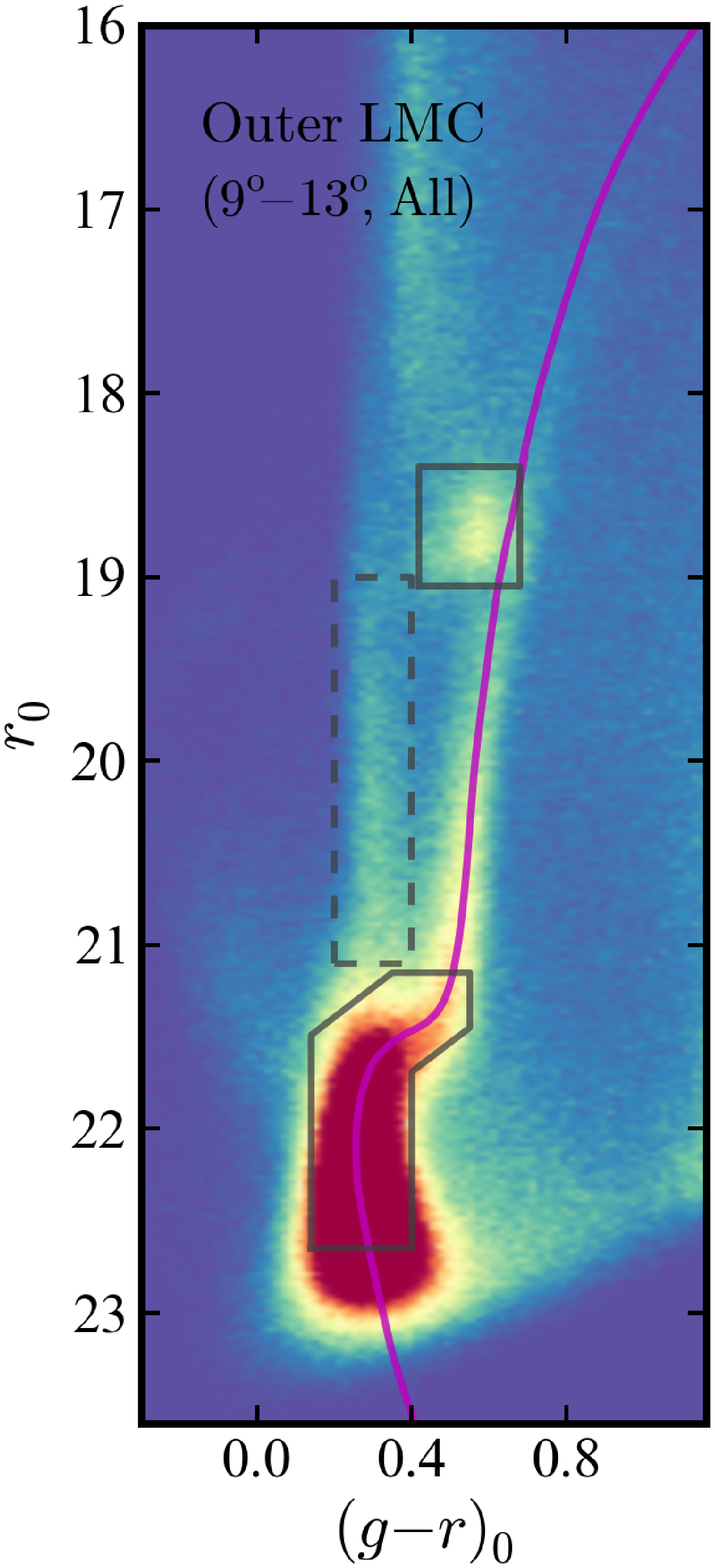}
\hspace{0mm}
\includegraphics[height=82mm]{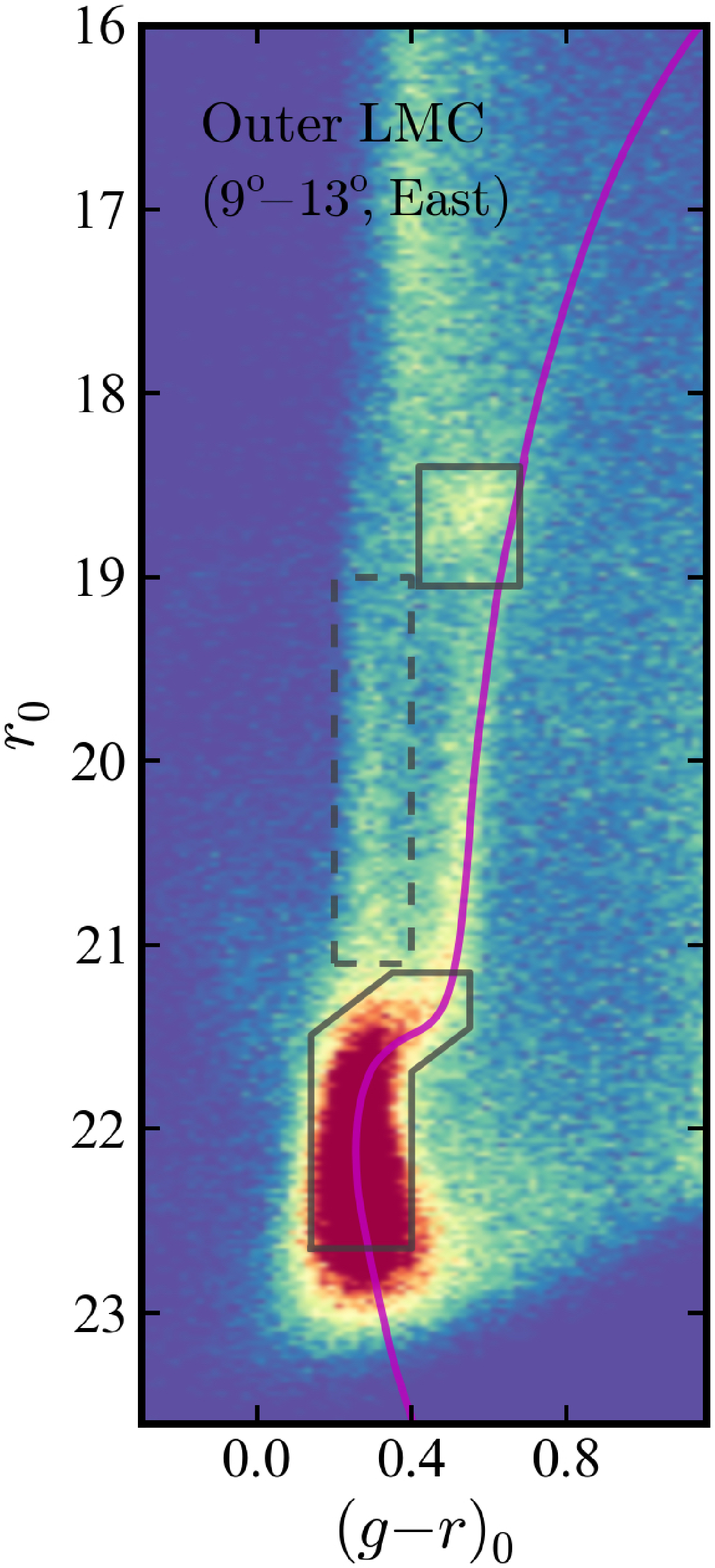}
\hspace{0mm}
\includegraphics[height=82mm]{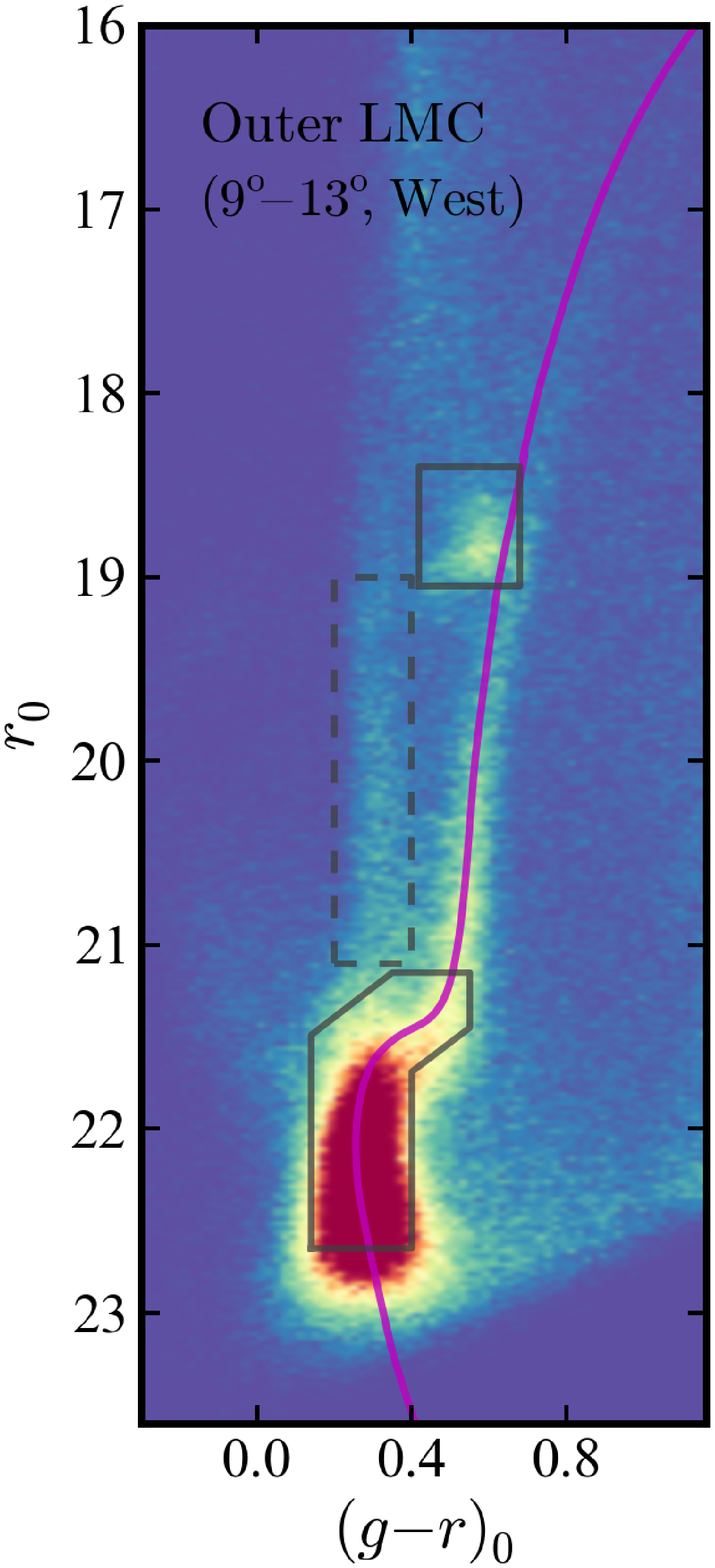}
\hspace{0mm}
\includegraphics[height=82mm]{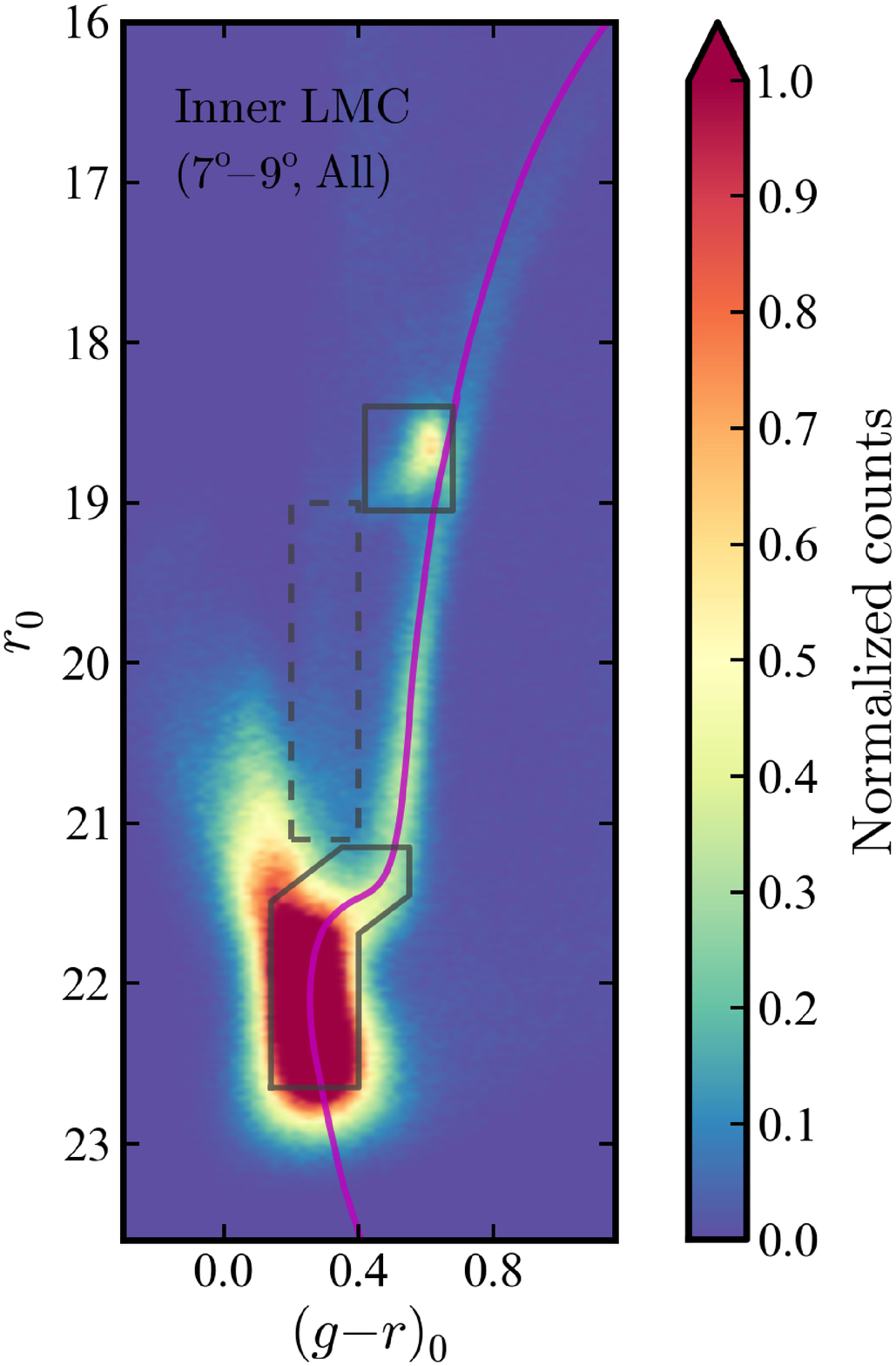}
\end{center}
\caption{Hess diagrams for stars in the northern outskirts of the LMC. {\bf Left:} Stars at all position angles covered by the 
year one DES footprint, and in the radial range $9\degr-13\degr$. {\bf Centre left:} Stars located to the far east of the footprint, 
at position angles $\ge 25\degr$, and in the radial range $9\degr-13\degr$. {\bf Centre right:} Stars located to the far west of the 
footprint, at position angles $\le -55\degr$, and in the radial range $9\degr-13\degr$. {\bf Right:} Stars at all position angles
covered by the year one DES footprint, and in the radial range $7\degr-9\degr$. In all four panels the magenta line denotes an 
isochrone from the Dartmouth Stellar Evolution Database \citep{dotter:08} with age $11$ Gyr, $[$Fe$/$H$] = -1.1$, and shifted
to a distance modulus of $18.35$. Note the clear line-of-sight distance variation between the two central CMDs relative to this 
fiducial line. We mark our selection box for old LMC main sequence turn-off stars using a solid line, together with a box 
indicating the location of the red clump belonging to this population. We mark our selection box for exploring contamination
from the Galactic foreground (Section \ref{ss:effects}) with a dashed line. In each panel the colour-map represents the stellar 
density in $0.02\times0.02$ mag pixels, smoothed with a Gaussian kernel of $\sigma = 0.02$ mag, and normalised to $50\%$ 
of the maximum pixel value. This normalisation has the effect of saturating regions of the CMD and introducing artificial 
fluctuations in the apparent level of non-LMC populations, but increases the contrast in low-density regions without moving 
to a non-linear scale.
\label{f:cmd}}
\end{figure*}

In the leftmost panel of Figure \ref{f:cmd} we show a Hess diagram for all stars in our catalogue lying in the range 
$9\degr-13\degr$ from the centre of the LMC.  Most prominent is the old stellar population of the outer disk, which is 
well described by an isochrone from the Dartmouth Stellar Evolution Database \citep{dotter:08} of age $11$\ Gyr,
$[$Fe$/$H$] = -1.1$, and $[\alpha/$Fe$] = 0.0$. Our photometry 
reaches $\approx 1$\ magnitude below the main-sequence turn-off belonging to this population. A distinct red clump 
is visible (marked with a box), but only a very faint blue horizontal branch, indicating that the peak of the metallicity distribution 
function sits around $[$Fe$/$H$] \approx -1$ or higher. This is consistent with previous investigations of outer LMC fields 
\citep[e.g.,][]{saha:10,carrera:11}.  A trace younger population may extend to the blue but this is completely sub-dominant.
The rightmost panel of Figure \ref{f:cmd}, shows a Hess diagram for all stars in the range $7\degr-9\degr$ from the centre 
of the LMC, and a substantial younger component is clearly present. This demonstrates that such populations are almost
completely confined to a radii smaller than $\approx 9\degr$, at least to the north of the LMC -- again consistent with
previous studies.  For example, \citet{balbinot:15} observed that populations younger than $\sim 4$ 
Gyr were truncated at $\approx 9\degr$ or $10\degr$ from the LMC centre in their survey fields to the north-west.

Returning to the leftmost panel of Figure \ref{f:cmd}, the distance modulus required to shift the fiducial isochrone onto the 
ridgeline of the dominant old stellar population is $\approx 18.35$.  Taken at face value, this suggests that the region of the 
LMC imaged in the year one DES footprint mostly lies somewhat closer than does the galaxy centre, which has a distance 
modulus of $\approx 18.5$ \citep[throughout this work we assume the distance to the LMC centre to be $D_0 = 49.9$\ kpc,][]{degrijs:14}.  
It is important to note, however, that our measurement is subject to systematic uncertainty of at least $0.1$ mag, due to the fact that 
we do not know precisely the age, metallicity, and $\alpha$-element abundance of the stellar population(s) in the CMD, and because 
distances derived using isochrones can be rather model-dependent. 

Another notable feature of our CMD is that both the sub-giant branch and the red clump of the dominant old population 
are vertically quite broad.  While this could be due to the presence of stars with a mix of metallicities and (old) ages, it might 
also reflect a spread in the distances to LMC stars in our catalogue.  Indeed, we find the latter possibility to be the most likely
explanation -- in the two central panels of Figure \ref{f:cmd} we plot Hess diagrams for stars lying in the radial range 
$9\degr-13\degr$, but at position angles $\ge 25\degr$ (centre) and $\le -55\degr$ (right) -- i.e., to the extreme
east and west of the imaged region. A significant distance gradient is apparent, in the sense that stars further to the east
are closer than those to the west. The sub-giant branch in the middle panel lies $\approx 0.15$\ magnitudes above the
fidicual isochrone, while that in the right-hand panel sits $\approx 0.1$ mag below the fiducial line; there is also a striking
difference in the red clump levels between the two CMDs.  These {\it relative} measurements are much more robust than our
determination of the mean absolute distance modulus above. They are consistent, at least in a qualitative sense, with
the recent work of \citet{balbinot:15}, who examined the brightness of the red clump in populations lying within 
$\approx 9\degr$ north-west of the LMC centre, and also \citet{vdm:01a} who studied the brightness of AGB stars as a function 
of radius and azimuth out to $\approx 7\degr$. The distance gradient is a result of the LMC geometry, which is usually
assumed to follow that of an inclined disk with regions to the north and east tilted in front of the plane of the sky.

\begin{figure*}
\begin{center}
\includegraphics[width=135mm]{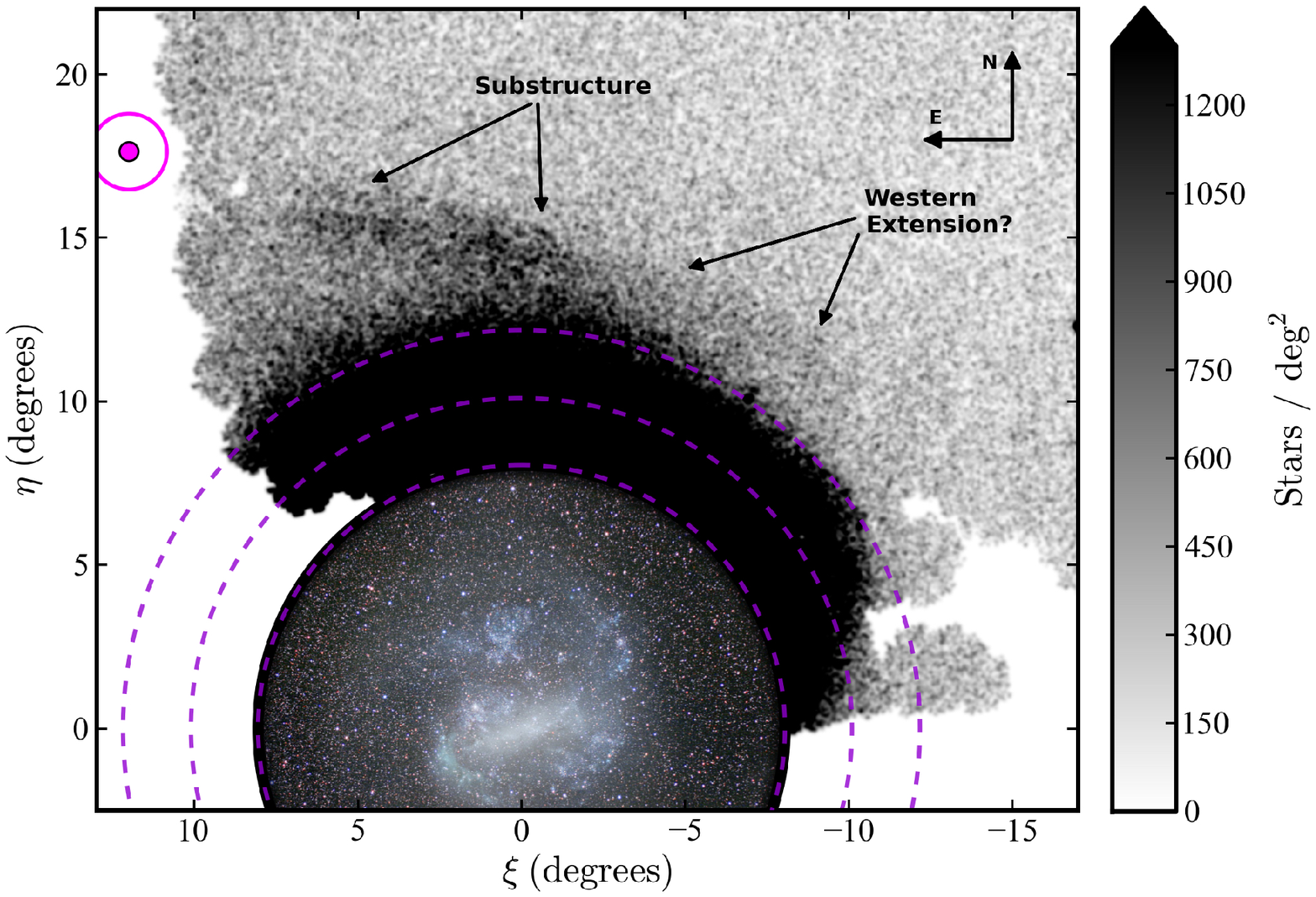}
\end{center}
\caption{Map showing the spatial density of stars consistent with being members of the LMC old main 
sequence turn-off population. These were selected from our photometric catalogue using the CMD box marked in
Figure \ref{f:cmd}. The colour-map has been set to saturate at a density of $1350$ stars deg$^{-2}$ -- this obscures
detail in the LMC outer disk, but enhances the visibility of low-density structures without moving to a non-linear
scale. The bin size is $2\arcmin \times 2\arcmin$; smoothing has been applied using a Gaussian kernel with 
$\sigma = 3\arcmin$. On this map, north is up and east is to the left as indicated; $(\xi,\,\eta)$ are coordinates in the 
tangent plane (gnomonic) projection centred on the LMC.  The three purple dashed circles mark angular separations of
$8\degr$, $10\degr$, and $12\degr$ from the LMC centre. Within $8\degr$ we display a wide-field optical image of 
the LMC (credit: Yuri Beletsky; used with permission) to help place the size of the periphery into perspective. 
Our new stream-like stellar substructure is clearly evident to the north of the outer LMC disk stretching in the direction 
of the Carina dwarf, which we mark with a magenta point and a surrounding circle that denotes an approximate 
radius of $1\degr$. The additional faint substructure to the west is also indicated.
\label{f:map}}
\end{figure*}
 
In the following analysis we aim to use main sequence turn-off stars belonging to the dominant old stellar population
seen in Figure \ref{f:cmd} to trace the outer structure of the LMC. However, the apparent variation in distance between 
different regions of the galaxy makes uniformly selecting these stars problematic because we do not know, {\it a priori}, 
the line-of-sight distance to any given point on the sky, and the stellar density is much too low to reliably measure this 
quantity with the necessary spatial resolution.  The simplest solution is to define a selection box on the combined 
CMD that is sufficiently broad to encompass the full vertical spread in the old population. This box is marked in Figure 
\ref{f:cmd}, and comes at the cost of a slightly higher level of non-LMC contaminants than we might have achieved 
with a more closely-tailored selection region\footnote{Note that the red clump is {\it not} included as part of our selection
region; the box marked in Figure \ref{f:cmd} is simply to help facilitate our discussion of this feature and its changing
level with position. Note also that while the red giant branch between the main sequence turn-off and the red clump is
visible, we do not include this region in our selection box in order to minimise contamination from Galactic foreground
stars -- see Section \ref{ss:effects}.}.

We set the faint limit of our selection box at a level that mitigates any potential issues due to the onset of (spatially variable) 
detection incompleteness at the level quoted in the previous section (see also Section \ref{ss:effects} below).
Assuming that the apparent distance gradient from east to west is smooth, holding this faint limit constant with position 
will introduce an artificial density gradient on the sky. That is, a patch of given stellar density will be observed to be rather
less dense in the west than in the east due to the CMD selection box in the west covering $\approx 0.25$\ magnitudes 
less of the old population main sequence than it does in the east. We will return to this issue later in the paper; for now 
we simply note that the size of this variation is of order $\sim 20$ per cent from the extreme eastern to western edges 
of the imaged LMC region.

\subsection{Stellar density map}
\label{ss:map}
In Figure \ref{f:map} we show a map of the spatial density of stars falling in our CMD selection box. This is a tangent-plane
(gnomonic) projection where the tangent point is located at the centre of the LMC \citep[which we assume to lie at 
$\alpha = 82.25\degr$, $\delta = -69.5\degr$ as in][]{vdm:01a}. The dashed purple lines mark galactocentric radii of 
$8\degr$, $10\degr$, and $12\degr$; within the central $8\degr$ we have superimposed a wide-field optical image of
the LMC to help place the periphery into perspective. In the DES density map the outer LMC disk dominates to a radius 
of $\approx 12\degr$ and possesses a mild but noticeable non-circular shape, as seen in previous studies at smaller
radial distances \citep[see e.g.,][]{vdm:01b}.  

Perhaps the most striking feature on the map is a distinct stream-like substructure stretching from $\approx 13.5\degr$ due
north of the LMC centre towards the eastern edge of the survey footprint. The apparent width of this feature on the
sky is $\approx 2\degr$.  Additional faint structure is visible at commensurate radii to the west; based on
the density map alone, it is not clear whether this constitutes an extension to the main body of the substructure or whether
it is unrelated (but see also Section \ref{ss:profile}).

The existence of this stream-like feature helps reconcile the results of several previous studies of the LMC outskirts:
\begin{enumerate}
\item{A number of authors, while investigating the Carina dwarf galaxy, have noted the presence of probable LMC 
stars in the foreground. These include giants and red clump stars observed spectroscopically by \citet{majewski:00} and 
\citet{munoz:06} that appear to constitute a kinematically cold structure with heliocentric radial velocity 
$\approx 332$\ km$\,$s$^{-1}$, as well as stars at the old main-sequence turn-off seen recently in a photometric study by 
\citet{mcmonigal:14}. The map presented by these latter authors shows the LMC populations to be mostly concentrated 
around $\sim 1\degr$ south of Carina, with a relatively sharp edge. In Figure \ref{f:map} we mark the position of the Carina
dwarf with a magenta point; the outer circle marks a distance of approximately $1\degr$ from the dwarf. Extrapolating the 
position and direction of our stream-like feature a short distance off the eastern edge of the DES footprint, we posit that 
these previous studies have very likely been detecting members of this substructure.}
\item{\citet{majewski:09} detected LMC giant stars spectroscopically to a radius of $\approx 19\degr$ to the north-east, 
north, and north-west, and to a radius of $22\degr$ in the direction of Carina.
On the other hand, \citet{saha:10} photometrically detected main sequence turn-off stars in fields to $16\degr$ due north 
of the LMC, but not in fields at $17\degr$ and $19\degr$. Our newly discovered substructure possesses a rather sharp 
drop off in density at $\sim15-16\degr$ due north of the LMC centre. This could explain the apparently conflicting results
of these two studies if the stellar density beyond this ``edge'' fell below the faint detection limit of \citet{saha:10} but not
\citet{majewski:09}. LMC populations have now been detected photometrically in this direction beyond $\sim15-16\degr$
\citep[see e.g., Section \ref{ss:profile} in the present work, and][]{belokurov:16}.}
\end{enumerate}
 
\begin{figure*}
\begin{center}
\includegraphics[width=85mm]{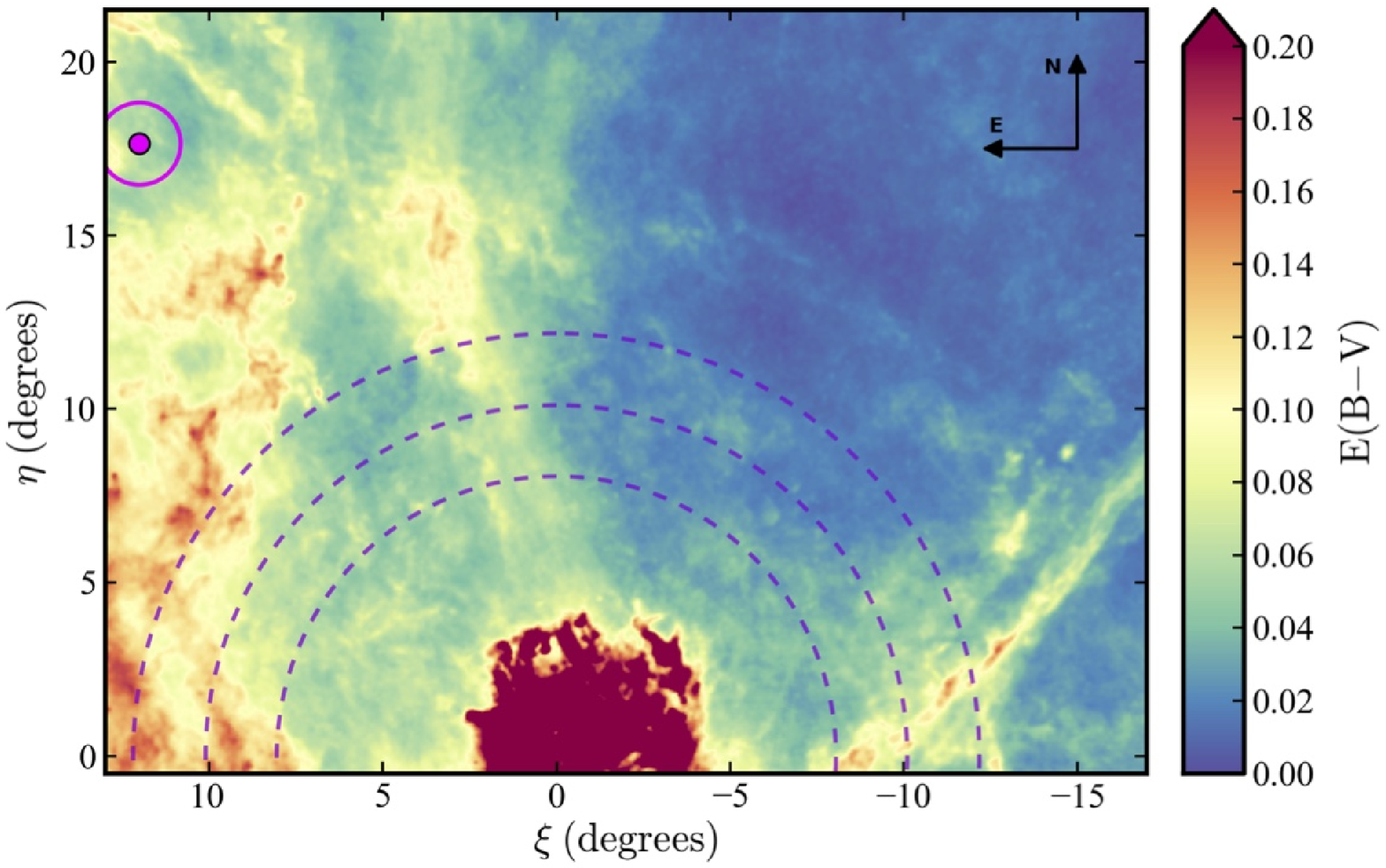}
\hspace{1mm}
\includegraphics[width=85mm]{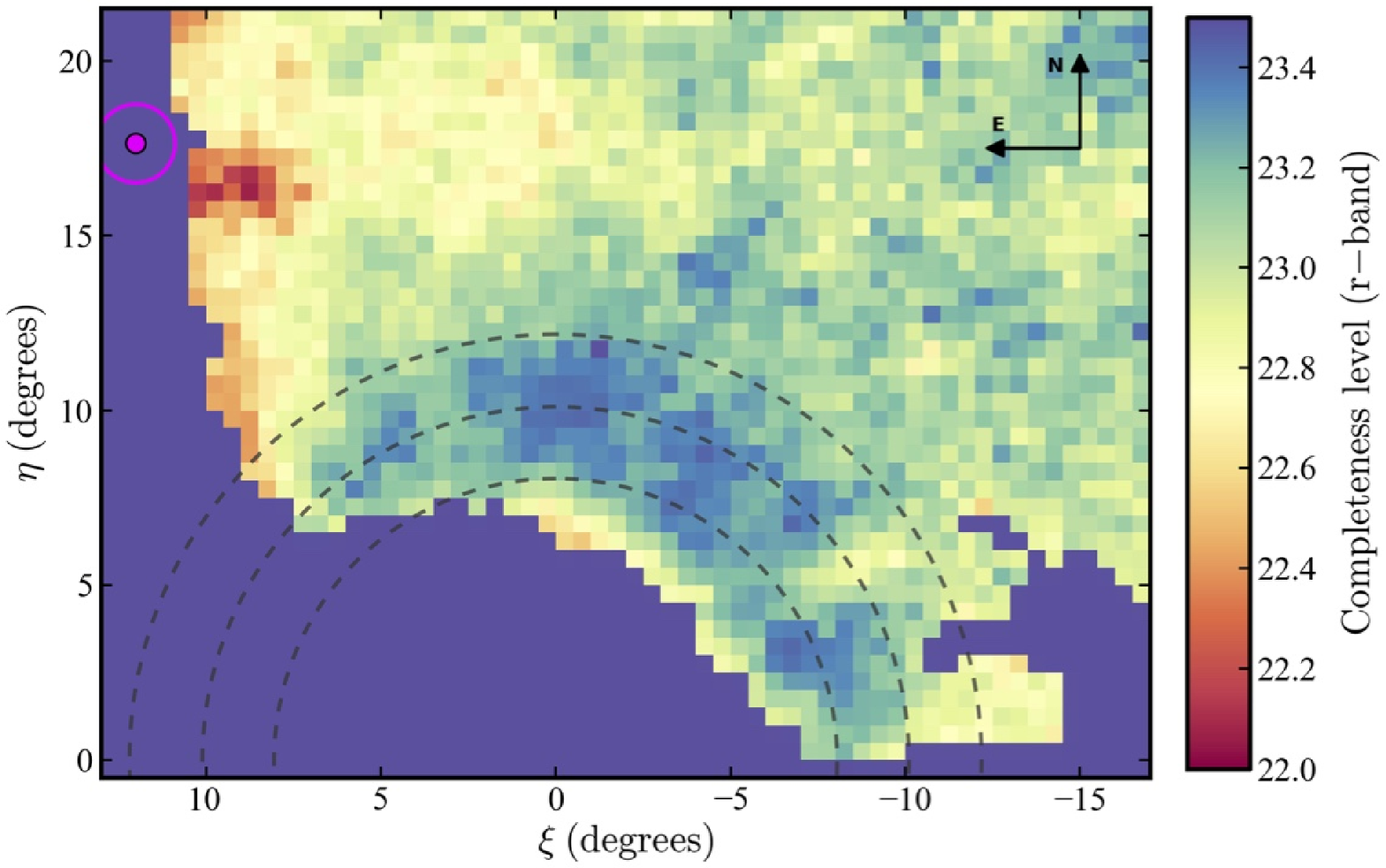}\\
\vspace{2mm}
\includegraphics[width=85mm]{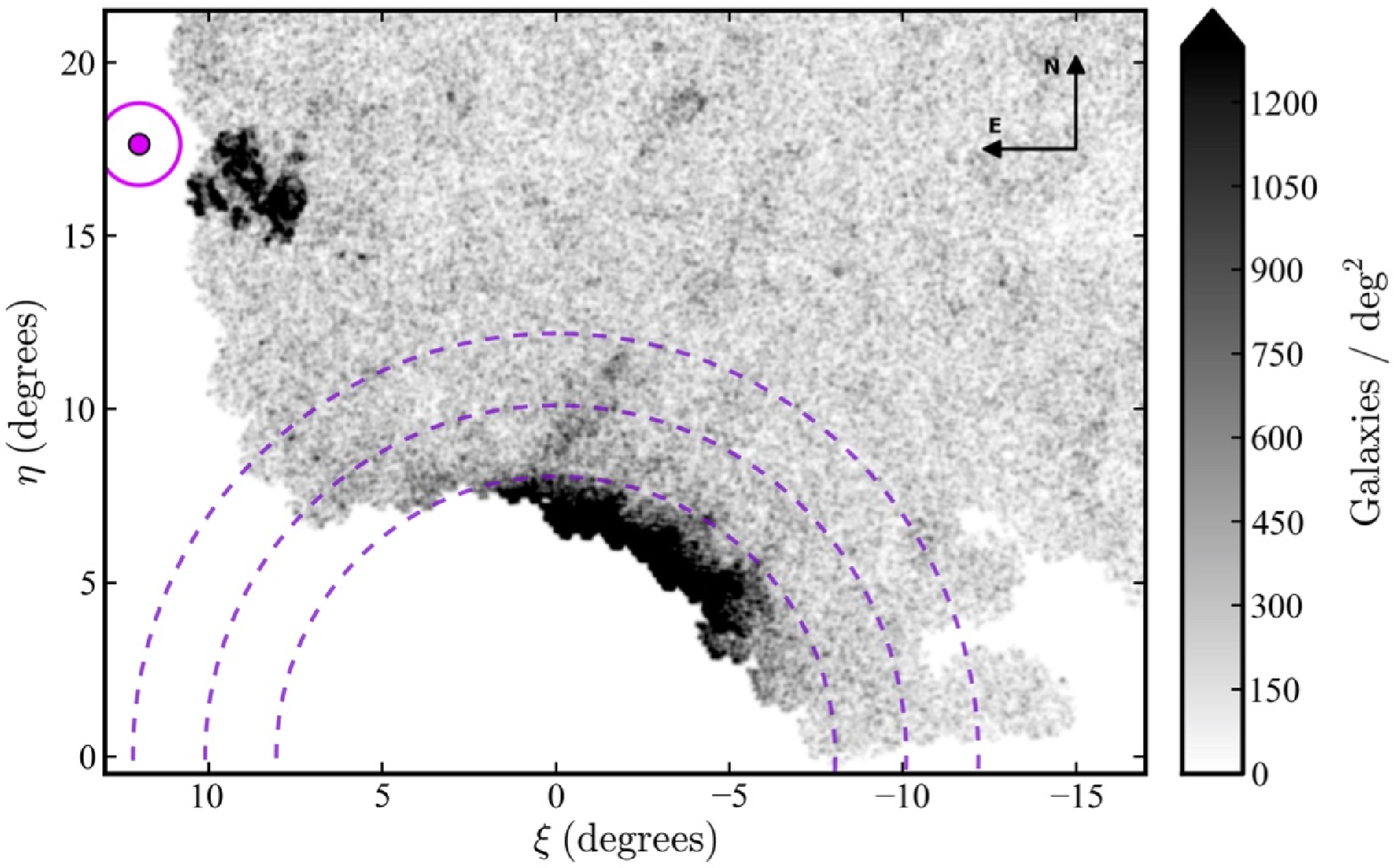}
\hspace{1mm}
\includegraphics[width=85mm]{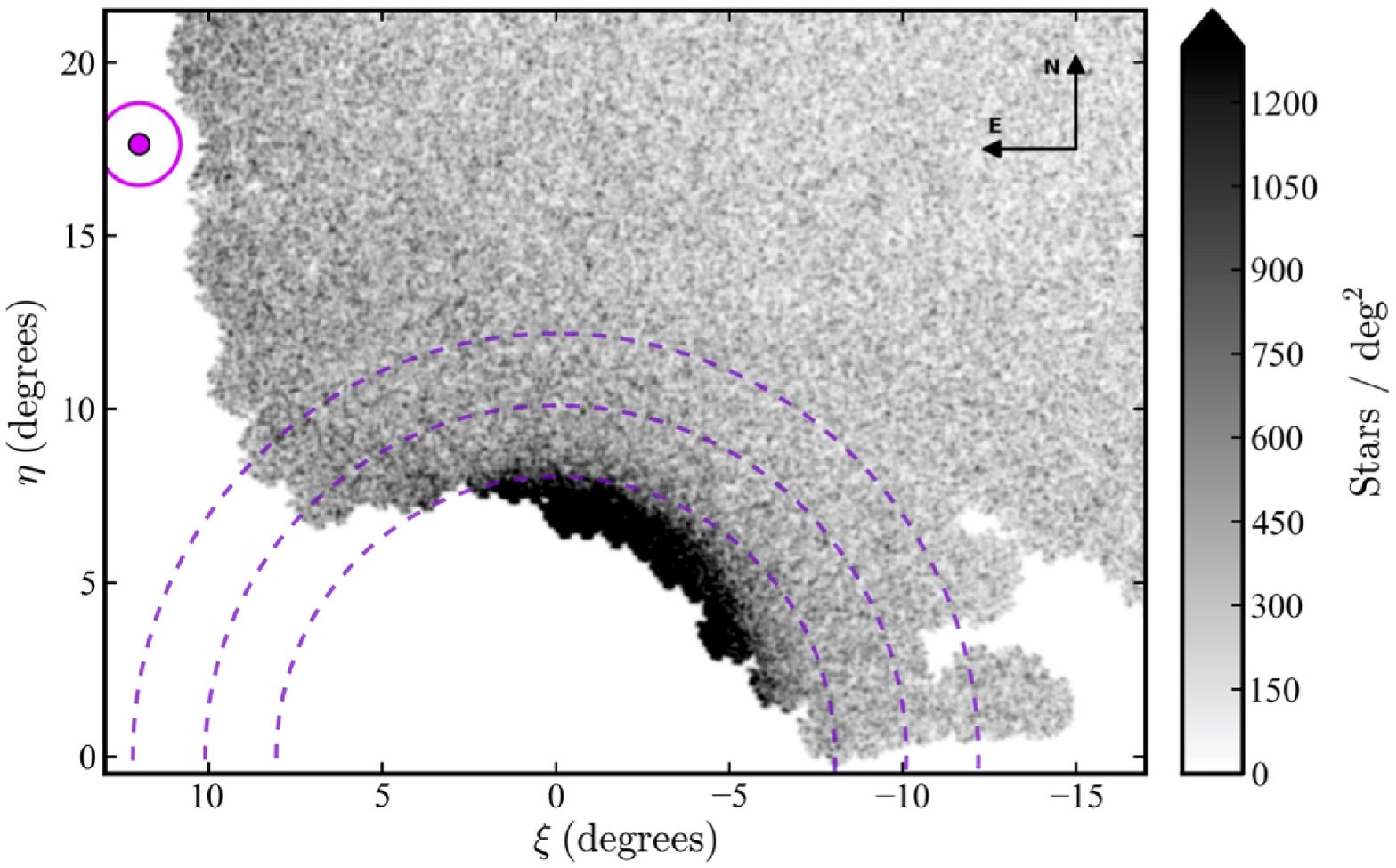}
\end{center}
\caption{Maps showing the variation in the following quantities across the region of interest: {\bf Upper left:} foreground
reddening, derived from the \citet{schlegel:98} dust map; {\bf Upper right:} detection incompleteness, calculated in 
$0.5\degr \times 0.5\degr$ boxes; {\bf Lower left:} the spatial density of objects classified as galaxies according to our
star-galaxy separation routine; and {\bf Lower right:} the spatial density of stars falling into the foreground selection box
defined on the CMDs in Figure \ref{f:cmd}.  For the latter two maps the colour-scaling is set to be equal to that
in Figure \ref{f:map} to facilitate easy comparison. In all four panels the three dashed circles denote angular separations of
$8\degr$, $10\degr$, and $12\degr$ from the LMC centre, and the Carina dwarf is marked with a magenta point.
\label{f:effmap}}
\end{figure*}

\subsection{Effects of reddening, incompleteness, and contamination}
\label{ss:effects}
It is important to check the impact of spatially variable foreground reddening, detection incompleteness, or contamination of 
the CMD selection box by non-LMC objects.  Each of these could, in an extreme case, give rise to an artefact that might mimic
a stream-like feature; more generally, these issues might alter the observed morphology or properties of the substructure.

The upper left panel in Figure \ref{f:effmap} shows the variation in $E(B-V)$ across the region of interest, from the \citet{schlegel:98}
dust map. As previously stated, in general the foreground reddening to the north of the LMC is mild, with a median value of
$E(B-V)\approx0.05$. Although the reddening increases in the east (in the direction of the Galactic plane), $E(B-V)$ is still always
below $\approx 0.14$ mag. Most importantly, there is no feature in the reddening map that is consistent with that of our 
substructure. Overall this suggests that variations in foreground reddening are not a significant factor for our analysis.

In the upper right panel in Figure \ref{f:effmap} we plot the spatial variation in the $r$-band magnitude where our catalogue starts
to become significantly affected by incompleteness. We define this in $0.5\degr \times 0.5\degr$ bins by locating the
turn-over in the stellar luminosity function and fitting a curve to determine the level at which the number counts fall to $75\%$
of this peak value. The intrinsic bin-to-bin scatter in this calculation is roughly $\pm 0.1$ mag. In general the completeness
level is relatively consistent across most of the region of interest, although there is a systematic move to brighter magnitudes
in the east.  This could be due to a general decrease in image quality or the effective integration time (i.e., fewer available epochs) 
near the edge of the DES year one survey footprint. There is also a noticeable move to brighter magnitudes inside $\sim 8\degr$ 
of the LMC centre due to
crowding. Near Carina there is a small patch where the completeness level moves to much brighter magnitudes -- this is a
region of poor data quality (see below). Importantly, apart from this small patch, variations in the completeness level always
occur at substantially fainter magnitudes than the cut-off level in our CMD selection box.  This is by design, to ensure that
spatially variable incompleteness does not affect our analysis. The bad patch of data falls squarely on the eastern end of our
substructure, and may explain why, in Figure \ref{f:map}, the feature appears to peter out towards the edge of the survey area.

Despite our CMD selection box being optimised to identify LMC main sequence turn-off stars, there will inevitably be some level
of contamination by non-LMC objects. The two primary sources of contamination are unresolved background galaxies, 
and stars belonging to the Galactic foreground. To investigate the former, we constructed a density map of objects classified
as galaxies rather than stars according to the star-galaxy separation routine described in Section \ref{s:data} 
\citep[see also][]{koposov:15}. While these objects are of course not {\it unresolved} galaxies, 
the expectation is that they will behave in a similar fashion.  This density map is shown in the lower left panel in Figure 
\ref{f:effmap}. In general, the distribution 
of galaxies is relatively uniform across the survey area, with a few filamentary regions of excess. The substantial excess 
towards the central regions of the LMC is due to crowding and the consequent misclassification of stars as galaxies. The patch
of bad data near Carina reappears in this map -- poor image quality in this region results in both the misclassification of stars
as galaxies, and the previously-observed bright completeness limit. 

Although there are a variety of Galactic stellar populations in the foreground, not all of these will contaminate our LMC selection
box on the CMD. In Figure \ref{f:cmd} we mark a second selection box (using a dashed line), that is designed to pick out old 
main sequence turn-off stars in the magnitude range $21 \la r \la 19$ -- i.e., at distances of $\sim 10-30$ kpc. The main sequences 
associated with these turn-off populations cut back towards the red and across the upper part of our LMC selection box; therefore, 
they serve as a good proxy for exploring how the foreground contamination should vary with position. We show a density map
for stars falling in this secondary selection box in the lower right panel in Figure \ref{f:effmap}.  Unsurprisingly, since most of
the stars in the distance range under consideration will belong to the Galactic halo, the distribution is very smooth. There is only
a very mild gradient towards the east, approaching the Galactic plane, presumably due to the nearest contaminants. In general
the expected level of contamination from foreground populations is only a small fraction of that of LMC stars, even in the 
outskirts of the LMC disk and including our new stellar substructure. We note that this analysis does not include main sequence
turn-off stars at distances beyond $\approx 30$ kpc, which will also fall into the LMC selection box; however in these regions
the Galactic halo is expected to have a negligible surface density even compared to the foreground map plotted in Figure \ref{f:effmap}.
Finally, we note an interesting apparent overdensity of foreground stars towards the centre of the LMC -- this excess is caused
by young LMC populations, that are present only within $\approx 8-9\degr$ of the centre, cutting across the foreground selection
box on the CMD (see the rightmost panel in Figure \ref{f:cmd}).

In summary, we have shown that spatially variable foreground reddening, detection incompleteness, and contamination of 
the CMD selection box by non-LMC objects cannot explain the presence of the stream-like feature observed in Figure 
\ref{f:map}, and it is thus reasonable to conclude that this is a genuine low surface-brightness substructure in the outskirts
of the LMC.

\subsection{Properties of the stream-like feature}
\begin{figure}
\begin{center}
\includegraphics[width=\columnwidth]{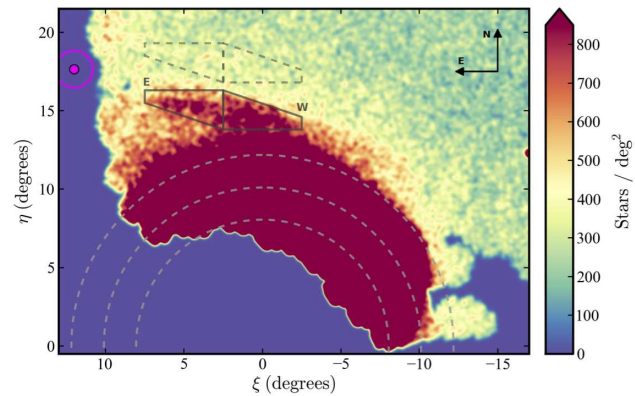}
\end{center}
\caption{A second realisation of our stellar density map, with the colour-scale set to saturate at $900$ stars deg$^{-2}$
and an increased degree of smoothing (using a Gaussian kernel of width $\sigma = 8\arcmin$). The selection box
used to explore the properties of the stellar populations in the stream-like substructure is marked with a solid line, and 
the corresponding region used for the field subtraction is marked with a dashed line. As before, the dashed circles indicate 
angular separations of $8\degr$, $10\degr$, and $12\degr$ from the LMC centre.
\label{f:mapsmr}}
\end{figure}

Figure \ref{f:mapsmr} shows a second version of our stellar density map, tuned to enhance low surface brightness
features in the LMC periphery. The faint structure in the outskirts of the disk 
to the west is clearly visible, as is some degree of structure to the south of the main stream-like feature stretching to
the east. In order to explore the properties of the stellar populations belonging to this feature, we define a selection 
box around its densest section.  As marked in Figure \ref{f:mapsmr}, this box is split into two equal regions -- one 
covering the eastern half of the substructure and one covering the western half. 

\begin{figure*}
\begin{center}
\includegraphics[height=83mm]{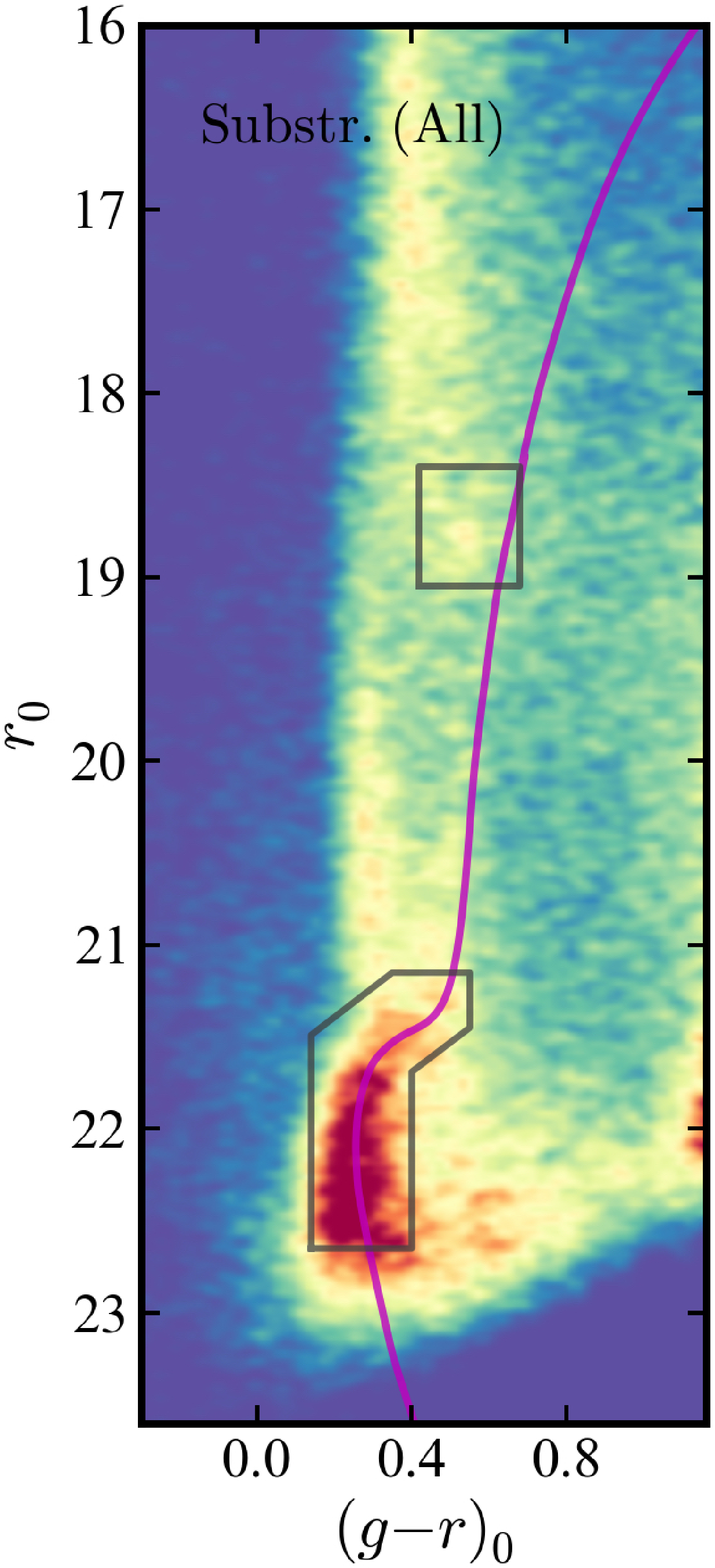}
\hspace{0mm}
\includegraphics[height=83mm]{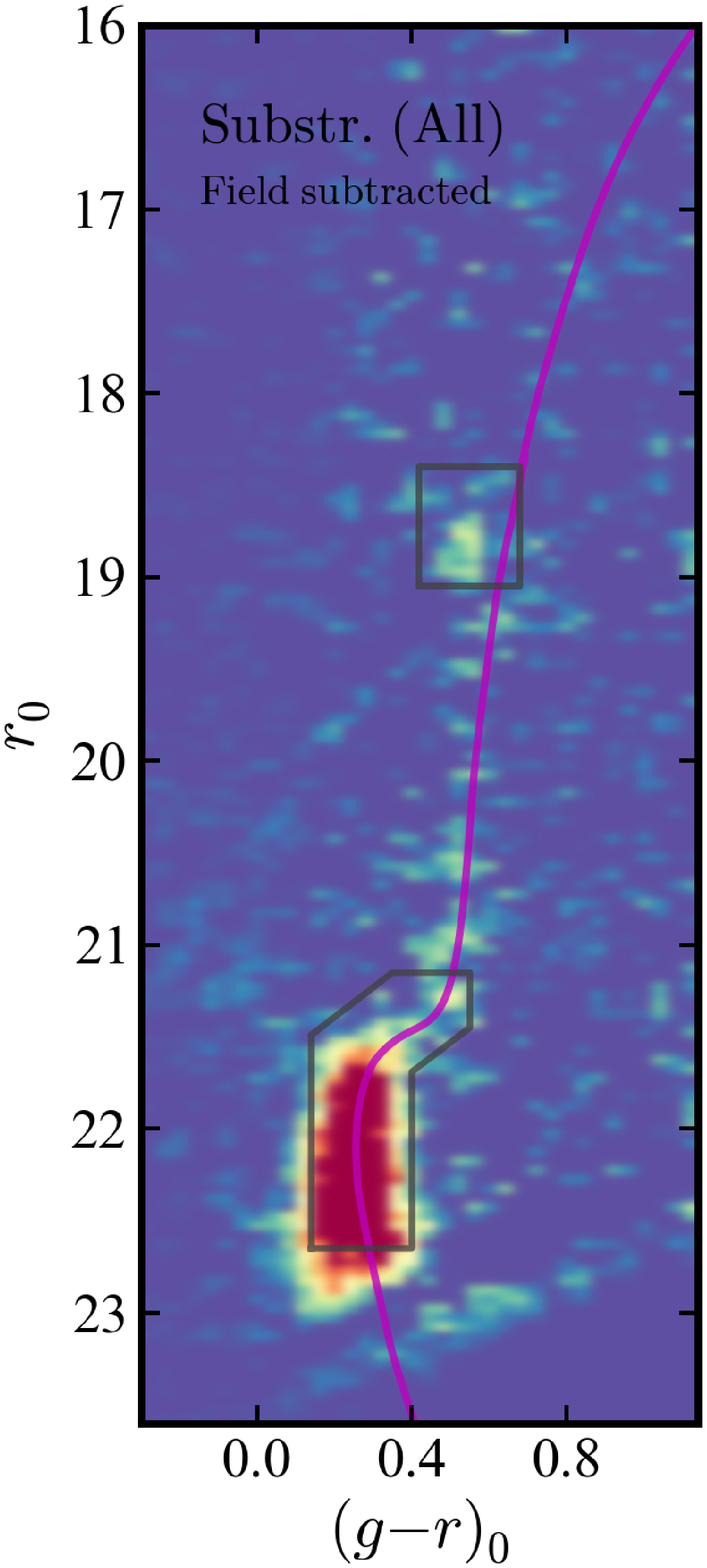}
\hspace{0mm}
\includegraphics[height=83mm]{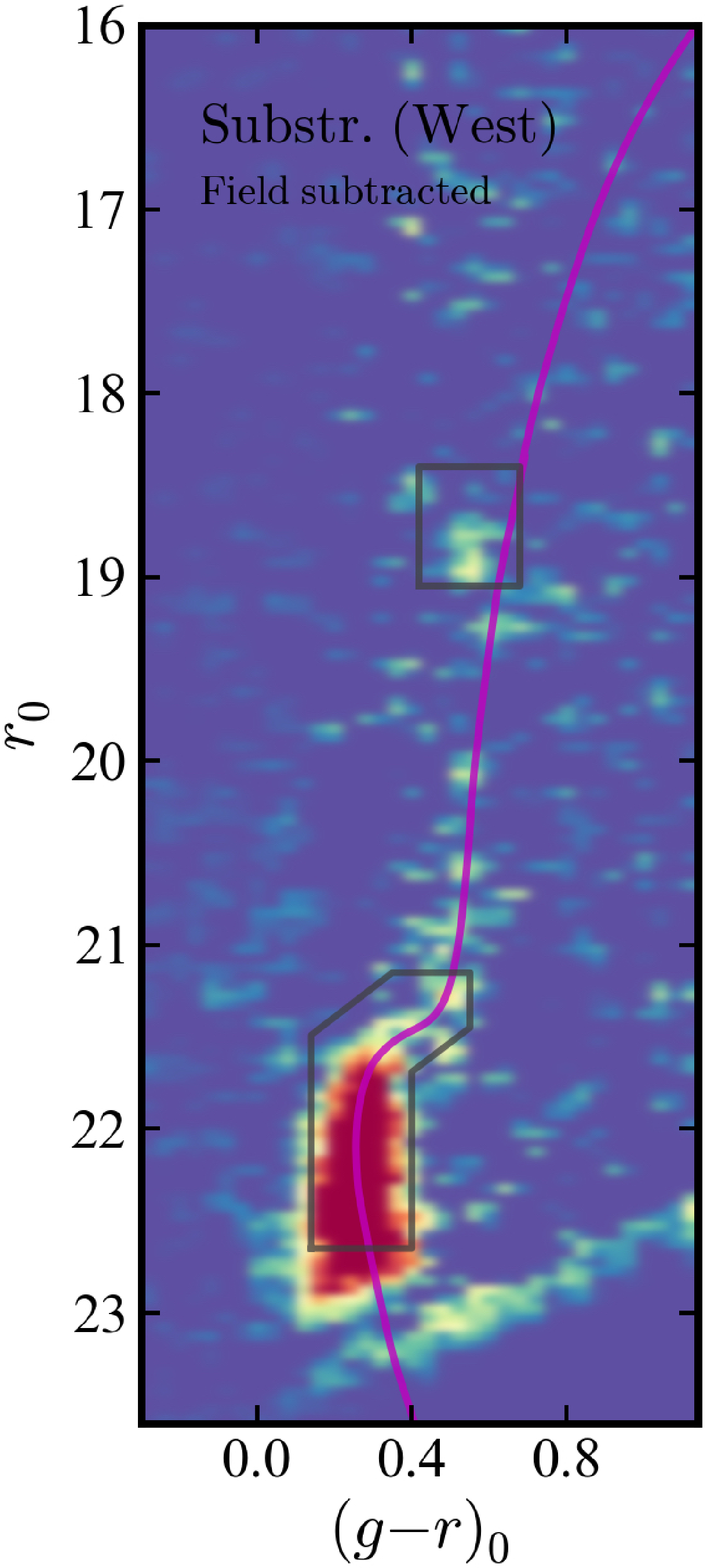}
\hspace{0mm}
\includegraphics[height=83mm]{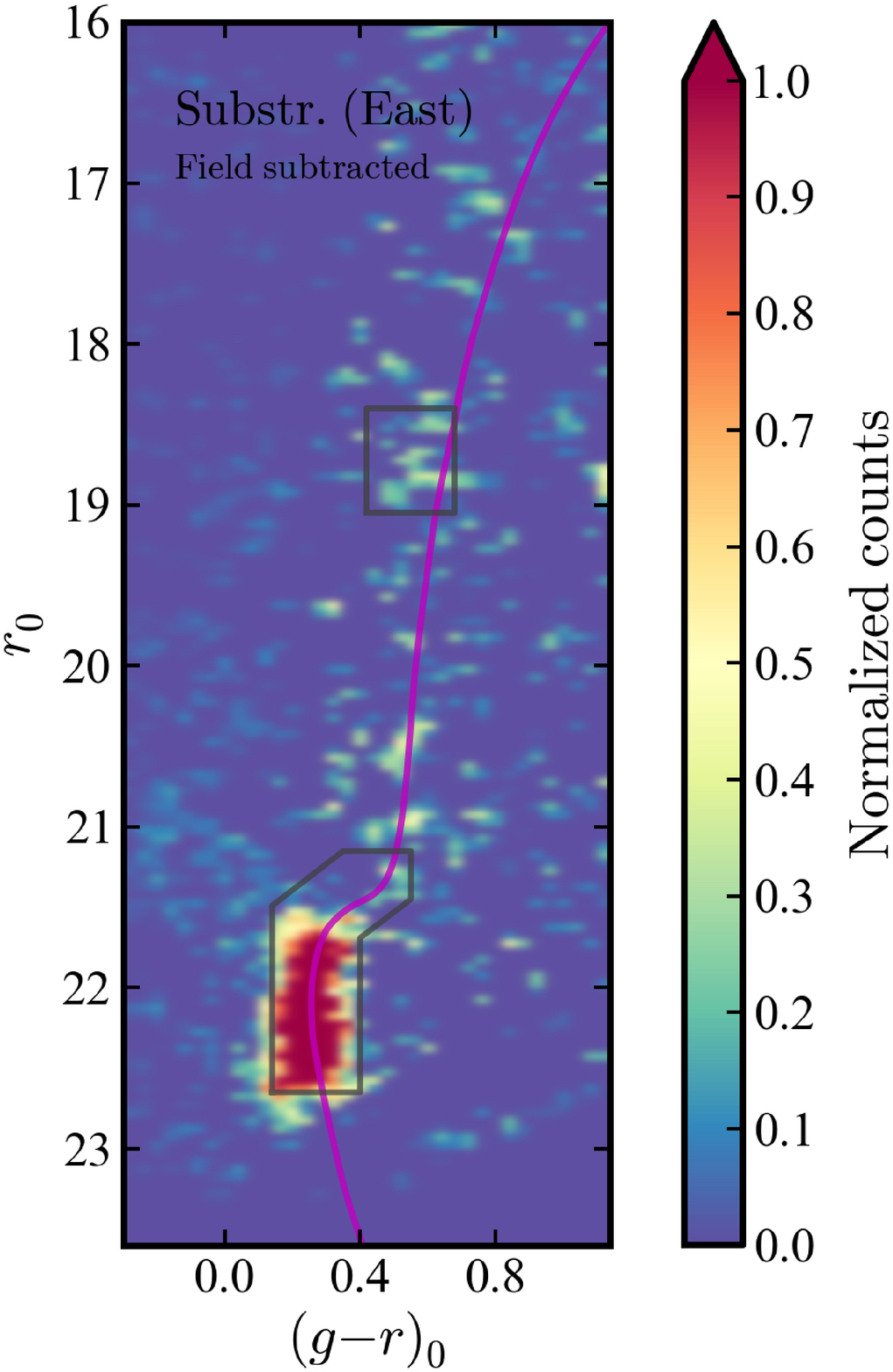}
\end{center}
\caption{Hess diagrams illustrating the properties of stars in the stream-like overdensity. {\bf Far left:} Stars lying in both halves 
of the selection box marked on the map in Figure \ref{f:mapsmr}. {\bf Centre left:} Field subtracted version of this Hess diagram.
{\bf Centre right:} Field subtracted Hess diagram for stars lying in the western half of the selection box. {\bf Far right:}
Field subtracted Hess diagram for stars lying in the eastern half of the selection box. In the left-most panel the CMD is
binned into $0.02\times 0.02$ mag pixels and smoothed with a Gaussian kernel of $\sigma = 0.02$ mag. In the other
three panels the CMDs are binned into $0.04\times0.04$ mag pixels and smoothed with a Gaussian kernel of 
$\sigma = 0.05$ mag. In all four panels the colour-scale is set to saturate at $40\%$ of the maximum pixel value, and
the same isochrone fiducial from Figure \ref{f:cmd} is marked with a magenta line. The same main sequence turn-off
and red clump selection boxes as in Figure \ref{f:cmd} are also plotted.
\label{f:streamcmd}}
\end{figure*}

In the left-hand panel of Figure \ref{f:streamcmd} we display a Hess diagram showing all stars falling within the selection 
box (i.e., within both halves). Although the foreground contamination is relatively severe, the signature of the substructure 
is clearly evident on this CMD -- both an old main sequence turn-off and a red clump are visible. The fiducial isochrone
that we used in Figure \ref{f:cmd} appears to match the locus of substructure stars quite closely, suggesting that the stellar 
population(s) that constitute this feature are very similar to the dominant old stellar population seen in the outer LMC disk. 

To investigate this further, we define a second box immediately to the north of the selection box for the substructure,
that covers an equivalent area of sky and which we use to define a Hess diagram for the foreground populations
that contaminate our stream CMD\footnote{Note that is is likely that this box contains some LMC-like stars
(see Section \ref{ss:profile}); however, these have a much lower spatial density than the substructure populations and
so are of no consequence here.}. We smooth both the ``substructure'' and ``field'' Hess diagrams with a Gaussian kernel
of $\sigma = 0.05$\ mag and then subtract the field distribution from that for the substructure via a straight grid point by
grid point differencing.  The resulting field-subtracted Hess diagram is displayed in the centre-left panel in Figure 
\ref{f:streamcmd}. The fiducial isochrone clearly provides an excellent description of the substructure populations; 
indeed even the distance of these populations appears to match quite closely to that of the isochrone. 
The red clump is clearly visible and corroborates this conclusion.

We repeated this procedure for stars in the western and eastern halves of the substructure selection box, to obtain the
Hess differences plotted in the centre-right and right-hand panels of Figure \ref{f:streamcmd}.  No significant 
variations in distance are evident from west to east along the feature, although we note that the CMD for the 
eastern half is very noisy due to the increasing level of foreground contamination in this direction.  This result
is consistent with the measurements of \citet{mcmonigal:14}, who found that the LMC stars they observed in front 
of the Carina dwarf lie $\approx 46$\ kpc away (i.e., at a distance modulus $\sim 18.3$).

The arc length along the substructure from $13.5\degr$ due north of the LMC centre to $1\degr$ due south of the 
Carina dwarf is $\approx 12.5\degr$ on the sky, corresponding to a physical length of $\sim 10$\ kpc. To estimate
a lower limit for its luminosity, we added up the total amount of light from stars in both parts of our 
selection box that also sit within the main sequence turn-off selection box on the CMD, and subtracted
the contribution of non-stream members using the field selection box. Ignoring any differences between DES $g,\,r$
and SDSS $g,\,r$, which are known to be small \citep[e.g.,][]{tucker:07}, we transformed from SDSS $g$ and $r$ 
magnitudes to the Johnson $V$-band using the relationship defined by Lupton on the SDSS DR12 web 
page\footnote{\url{http://www.sdss.org/dr12/algorithms/sdssUBVRITransform/}} to obtain $M_V = -5.7$. 

However, the main sequence turn-off stars sitting in our CMD selection box represent only a fraction of the total substructure 
population -- for example red giants, red clump stars, and faint main sequence stars are all excluded. To determine 
an approximate correction to account for light due to these missing populations we generated a luminosity function 
from the Darthmouth Stellar Evolution Database with the same age and metallicity as our fiducial isochrone, and based 
on an underlying \citet{kroupa:01} mass function. This revealed that our CMD selection box encompasses only 
$\approx 20$ per cent of the total light expected from a stellar population with these characteristics, implying 
a total luminosity of $M_V \approx -7.4$. This is still, of course, a lower limit to the true luminosity of the structure 
because our spatial selection box covers only its brightest portion.  We also note possible systematic uncertainties
of up to $\sim 0.5$ mag due to any spread in the characteristics of the substructure populations that we are unable to detect 
using the presently-available CMDs, as well as the inclusion of underlying disk populations that may or may not be appropriate 
depending on the origin of the feature (see Sections \ref{ss:profile} and \ref{s:discussion}). Assuming that $M_V \approx -7.4$ , 
the mean apparent $V$-band surface brightness of the substructure within the selection box is $\sim 31.8$\ mag arcsec$^{-2}$. 

\subsection{Structure of the outer LMC}
\label{ss:geom}
In order to try and obtain some insight into the origin of the stream-like overdensity, we now turn to a brief investigation 
of the structure of the outskirts of the LMC. Following previous studies \citep[e.g.,][]{weinberg:01,balbinot:15} we first attempt 
to model the geometry of this region 
using a circular exponential disk that is inclined to the plane of the sky. We adopt the formalism of \citet{vdm:01a} and
define coordinates $(\rho,\,\phi)$ on the celestial sphere for a given point with right ascension $\alpha$ and declination 
$\delta$, such that
\begin{align}
\cos\rho&=\cos\delta\cos\delta_0\cos(\alpha-\alpha_0)+\sin\delta\sin\delta_0\nonumber\\[6pt]
\tan\phi&=\frac{\cos\delta\sin\delta_0\cos(\alpha-\alpha_0)-\sin\delta\cos\delta_0}{\cos\delta\sin(\alpha-\alpha_0)}\,,
\end{align}
where $(\alpha_0,\,\delta_0)$ are the right ascension and declination of the centre of the LMC. The coordinate $\rho$ is the
angular distance between $(\alpha,\,\delta)$ and $(\alpha_0,\,\delta_0)$, while $\phi$ is the position angle of $(\alpha,\,\delta)$
with respect to $(\alpha_0,\,\delta_0)$ taken counterclockwise from the west\footnote{This is different from the usual astronomical 
convention where position angle is measured counterclockwise from the north.}.  As described earlier, we assume
$\alpha_0 = 82.25\degr$ and $\delta_0 = -69.5\degr$. This represents the approximate central point of the outer number 
density contours measured by \citet{vdm:01b} \citep[see also][]{vdm:01a} and as such is appropriate for the present problem. 

We next introduce a Cartesian coordinate system $(x,\,y)$ in the plane of the LMC disk\footnote{Note that this is the $(x^\prime,\,y^\prime)$
coordinate system defined by \citet{vdm:01a}.}; for simplicity we assume 
that the disk is infinitely thin. The disk plane is inclined with respect to the sky by an angle $i$ around an axis (the line of nodes)
that sits at position angle $\theta$, which is again measured counterclockwise from the west. The line of nodes represents the 
intersection of the LMC disk plane with the plane of the sky. The coordinates $(x,\,y)$ may be calculated from 
$(\rho,\,\phi)$ according to
\begin{align}
x&=\frac{D_0\cos i\sin\rho\cos(\phi-\theta)}{\cos i\cos\rho-\sin i\sin\rho\sin(\phi-\theta)}\nonumber\\[6pt]
y&=\frac{D_0\sin\rho\sin(\phi-\theta)}{\cos i\cos\rho-\sin i\sin\rho\sin(\phi-\theta)}\,,
\end{align}
where $D_0$ is the distance to the centre of the LMC, which, as noted previously, we assume to be $49.9$\ kpc. The line-of-sight
distance $D$ to any point on the LMC disk plane is given by
\begin{equation}
D=\frac{D_0\cos i}{\cos i\cos\rho-\sin i\sin\rho\sin(\phi-\theta)}\,.
\label{e:dist}
\end{equation}

In our model the surface density in the plane of the LMC disk varies as $\mu_0\exp(-r/r_s)$, where the radial distance 
$r^{2}=x^{2}+y^{2}$, the scale length is $r_s$, and the central surface density is $\mu_0$. Note that we do not include a background
contribution as the density of contaminants is much lower than the density of LMC stars over the region where we fit our
model (see below, and Section \ref{ss:profile}). We divide the 
tangent plane $(\xi,\,\eta)$ into a grid of $10\arcmin\times 10\arcmin$ cells and compare the observed number of stars 
per cell to that predicted by the model. In doing so, there are three additional factors we need to take into account.
First, the gnomonic projection $(\alpha,\,\delta)\rightarrow(\xi,\,\eta)$ does not conserve area. The distortion is approximately 
proportional to $1/(\cos\rho)^3$, such that a constant cell size on the tangent plane corresponds to an increasingly smaller area 
on the celestial sphere as the separation $\rho$ from $(\alpha_0,\,\delta_0)$ becomes larger; we use this to correct the predicted star counts.
Second, we must account for the fact that the line-of-sight distance to the inclined LMC plane varies across the sky.  
A cell for which the distance is shorter spans a smaller physical area on the plane than does a cell for which the distance is larger; 
the necessary scaling of counts per cell thus varies as $D^2$. 

Finally, as previously described in Section \ref{ss:cmds}, the line-of-sight distance variation also tends to move stars in and 
out of our fixed CMD selection box, which can lead to artificial density gradients across the sky. To attempt to incorporate this
into our model we define an empirical correction function $\lambda(\phi)$, as follows. We take all stars in the radial range 
$9\degr-13\degr$, as plotted in the left-hand panel of Figure \ref{f:cmd}, and split them by position angle into six equal-sized 
subgroups\footnote{This represents the finest partitioning for which we maintain adequate statistics on the Hess diagrams.} 
(the central-left and central-right panels in Figure 
\ref{f:cmd} represent the extreme eastern and western of these ensembles). For each subgroup we determine the vertical shift 
necessary to align the fiducial isochrone with the ridgeline on the Hess diagram, relative to the shift required for the eastern-most 
segment. We also determine the relative offsets in the red clump levels by measuring the position of the peak of a histogram in $r$ 
magnitude within the red clump selection box, and find strong consistency with the offsets derived using the fiducial isochrone. 
We reiterate that these {\it relative} distance measurements are much more robust to potential systematic uncertainties than absolute 
distances determined from isochrone fitting.

\begin{figure}
\begin{center}
\includegraphics[width=\columnwidth]{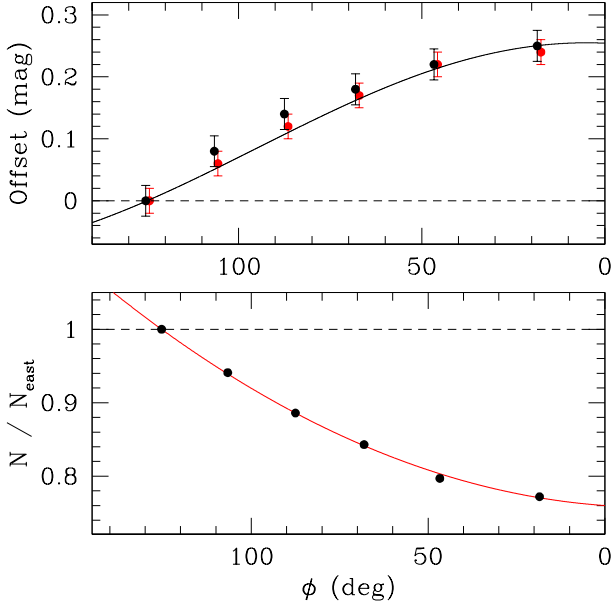}\\
\end{center}
\caption{{\bf Upper panel:} Relative distance measurements as a function of position angle $\phi$ for the six segments of the outer
LMC disk (radii $9\degr-13\degr$) described in the text. Recall that $\phi$ is measured counterclockwise from the west. All distances
are measured relative to the easternmost segment. Black points mark distances derived from aligning the fiducial isochrone with
the CMD ridgeline in the main sequence turn-off region, while red points mark distances derived from the red clump. There is good
agreement between the two sets of measurements. The black solid line shows a sinusoid with a peak-to-peak amplitude of $0.34$ mag
and a phase shift of $275\degr$. {\bf Lower panel:} Star counts for each of the six segments relative to the easternmost segment.
The variation is of order $20\%$ from east to west; the measurements are well described with a second-order polynomial (marked with
the red line).
\label{f:dgrad}}
\end{figure}

The upper panel of Figure \ref{f:dgrad} shows the results of our measurements (recall that the position angle $\phi$ is measured 
counterclockwise from the west). The line-of-sight distance appears to increase smoothly and monotonically from east to west,
and the gradient becomes shallower in the far west; the overall shape of the curve is consistent with being sinusoidal 
\citep[cf.][]{vdm:01a}. Because our data cover only about one-third of the outer LMC disk we are not able to constrain the phase
and amplitude of the sinusoid very precisely; however we overplot, relative to the distance zero-point at $\phi\approx 125\degr$, 
a representative fit which has a peak-to-peak amplitude of $0.34$ mag and a phase shift of $275\degr$.
Note that the stars in each azimuthal bin marked on the plot have a different mean angular distance from the LMC centre, due to 
the elliptical shape of the LMC disk on the sky.

To determine how the distance gradient affects our star counts, we took all stars found inside the CMD selection box for the 
eastern-most group, increased their magnitudes by the offset appropriate for each of the other segments in turn, and then counted how many 
still fell within the CMD selection box. The outcome is shown in the lower panel of Figure \ref{f:dgrad}.  As expected there is a 
substantial decrease in the number of stars falling in the selection box as a function of position angle.  The magnitude of the variation
is of order $20$ per cent from east to west across the survey footprint. Our measurements are well fit by a second-order 
polynomial\footnote{Although the locus of points appears to share a very similar shape to that for the relative distances in the panel 
above, such that the functional form may well be formally sinusoidal, a polynomial fit works well over the range in azimuth considered here.} 
of the form
\begin{equation}
\lambda(\phi) = 0.1255 \left(\frac{\phi}{100}\right)^2 + 0.3418\left(\frac{\phi}{100}\right) + 0.76\,.
\label{e:corr}
\end{equation}

Returning to our inclined exponential disk model, the overall number of stars predicted in a given cell $j$ is
\begin{equation}
m_j\left(\mu_0,\,r_s,\,\theta,\,i\right)=\mu_0 D_j^2(\cos\rho_j)^3 \lambda(\phi_j) \exp(-r_j/r_s)\,.
\end{equation}
The probability of finding the observed number of stars $n_j$ in this cell is given by a Poisson distribution
\begin{equation}
P_j\left(n_j\,|\,m_j\right)=\frac{m_j^{n_j} e^{-m_j}}{n_j!}\,,
\end{equation}
so that the logarithm of the likelihood of the data across all cells is
\begin{gather}
\mathcal{L}\left(n,\,\rho,\,\phi\,|\,\mu_0,\,r_s,\,\theta,\,i\right)=\ln\prod_{j}P_j\nonumber\\
=\sum_{j}\left[n_j\ln(m_j) - m_j - \ln(n_j!)\right]\,.
\end{gather}
The model includes implicit information about the distance to the stars in each cell, as this affects the observed density through both the 
$D^2$ and $\lambda(\phi)$ terms. Ideally we would also self-consistently include an explicit comparison of the measured distance to a given 
cell with that predicted by the model; however, as previously noted the stellar density is too low to allow distance measurements with useful
precision at the necessary spatial resolution.

\begin{table*}
\centering
\caption{Most-likely parameter values for our two inclined exponential disk models.}
\begin{tabular}{@{}lcccccccc}
\hline \hline
Model & Radial range & \hspace{1mm} & $\mu_0$ & $r_s$ & $\Theta$ & $i$ & $\psi$ & $e$ \\
 & & & (stars cell$^{-1}$ kpc$^{-2}$) & (kpc) & (deg) & (deg) & (deg) & \vspace{1mm}\\
\hline
Circular disk & $9\degr-13\degr$   & & $20.95\pm0.26$ & $1.50\pm0.01$ & $159.59\pm0.12$ & $33.14\pm0.09$ & $\ldots$ & $\ldots$ \vspace{2mm} \\
Elliptical disk & $9\degr-13\degr$ & & $12.80\pm0.25$ & $1.64\pm0.01$ & $(185.0)$ & $25.18\pm0.71$ & $34.68^{+2.31}_{-2.24}$ & $0.092\pm0.001$ \\
\hline
\label{t:ml}
\end{tabular}
\end{table*}
 
\begin{figure}
\begin{center}
\includegraphics[width=\columnwidth]{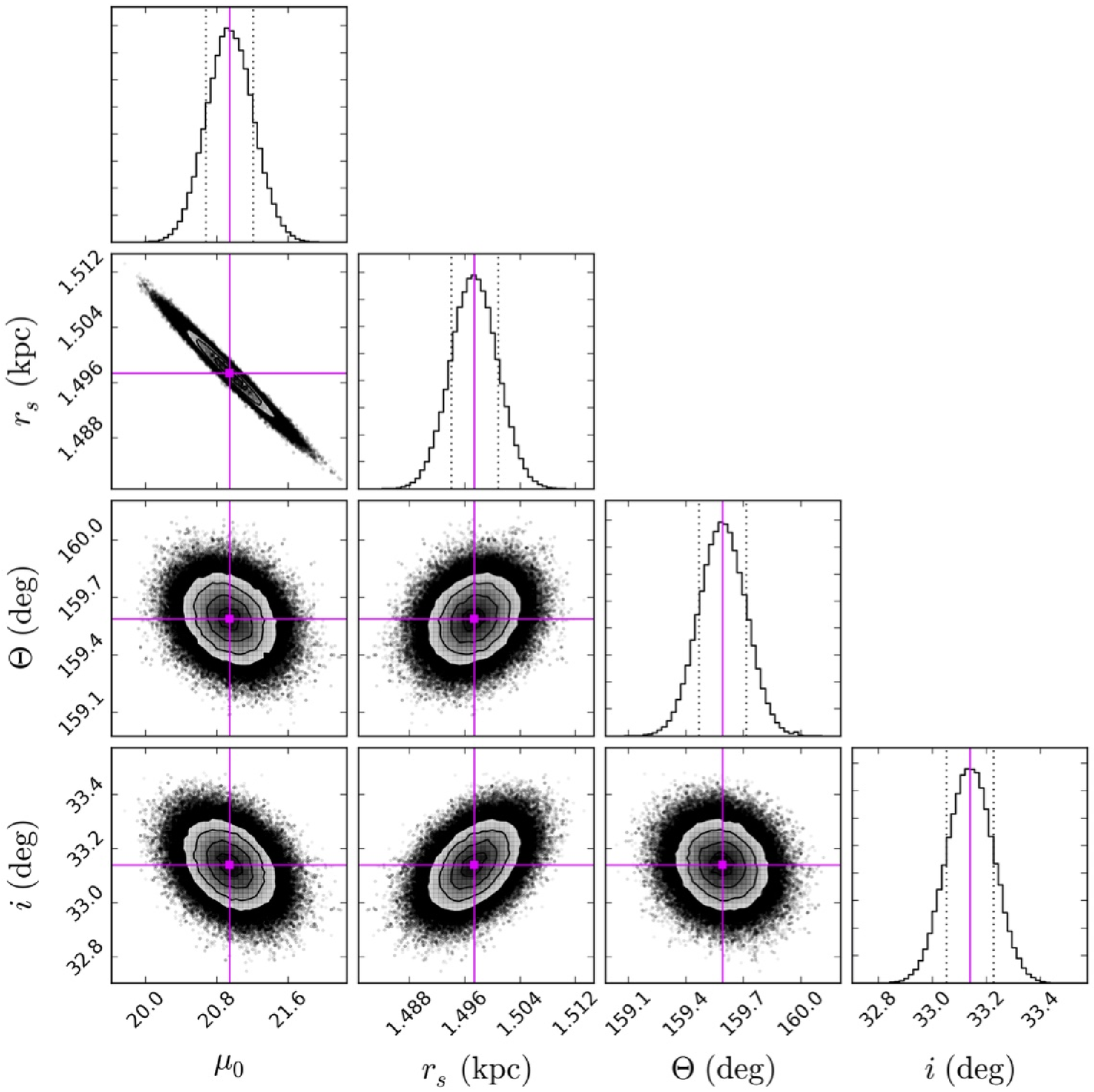}\\
\vspace{1mm}
\includegraphics[width=\columnwidth]{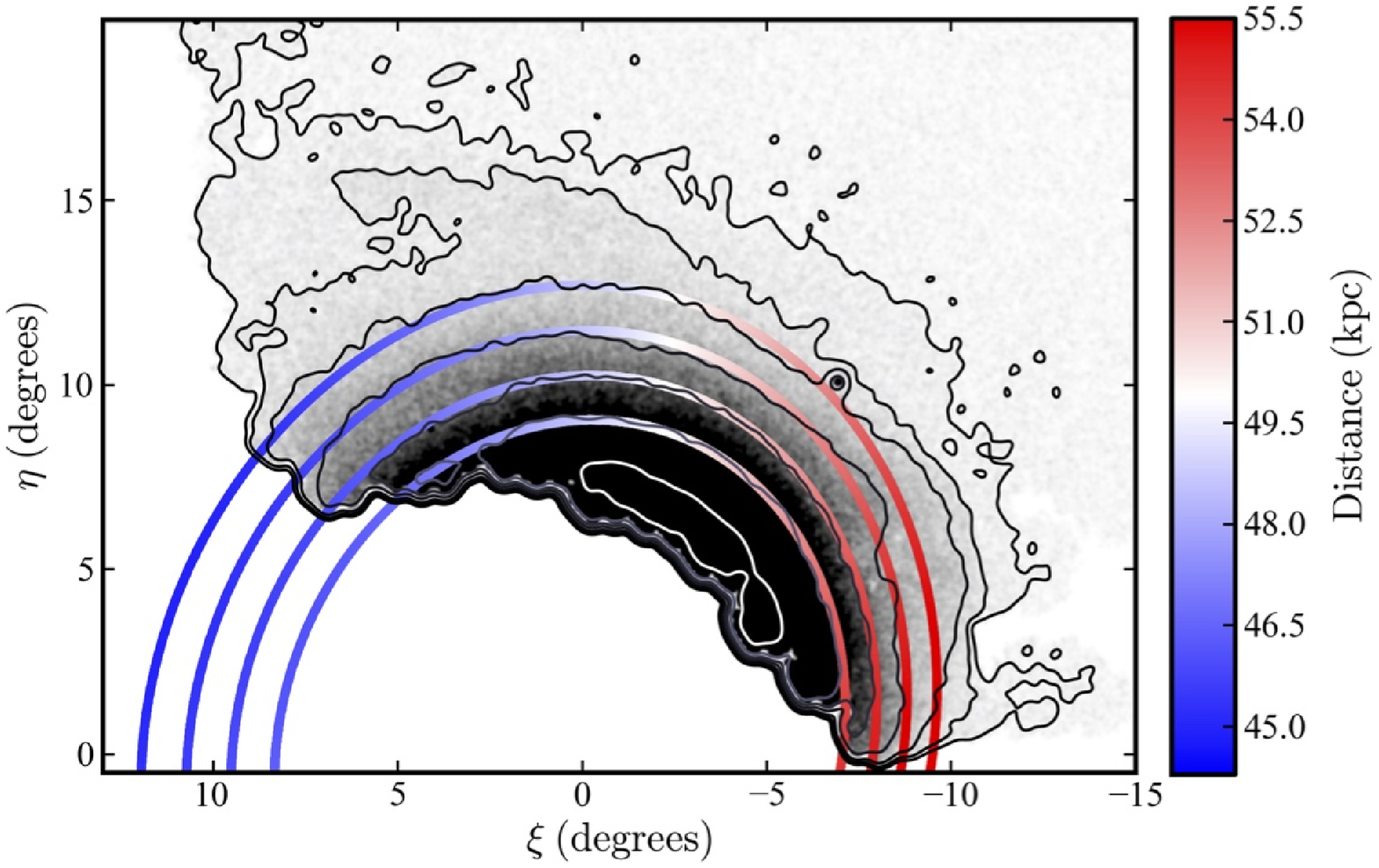}
\end{center}
\caption{{\bf Upper panels:} Marginalized one- and two-dimensional posterior probability distributions for the four
parameters in our inclined circular exponential disk model. These are the central density $\mu_0$, the scale length $r_s$, the 
position angle of the line of nodes $\Theta$ (measured counterclockwise from the north), and the inclination angle $i$.
The contours show the $1\sigma$, $2\sigma$, and $3\sigma$ confidence levels. {\bf Lower panel:} The best-fitting model
overplotted on our stellar density maps. Various isodensity contours are marked. The model is plotted at four different 
radii in the disk plane -- $8$, $9$, $10$, and $11$ kpc. The colour-scaling indicates the line-of-sight distance to each point
on the model, as indicated. In general there is good correspondence between the shape of the model and the isopleths.
North is up and east to the left. 
\label{f:mlcirc}}
\end{figure}

We use the Markov chain Monte Carlo sampler {\em emcee} \citep{fm:13} to infer the most likely set of model parameters
given our observations and the assumption of flat priors. We restrict $\mu_0$ and $r_s$ to be positive, and $0\degr\le i<90\degr$.
With this bound on the inclination, and our observation that LMC stars lying to the east of the survey footprint are closer than those lying 
to the west, it follows that $180\degr\le\theta<360\degr$. Our calculation considers all cells with $9\degr\le\rho\le13\degr$, where the 
inner bound is set to minimise contamination in our CMD selection box from young populations 
\citep[which may not be well described by an exponential disk model -- e.g.,][]{balbinot:15}, and the outer bound ensures we do not encroach 
on the region occupied by the substructure (which is a clear deviation from an exponential disk). We use an empirically-determined mask to 
excise the $\approx1\%$ of cells that intersect with small gaps in the spatial coverage, regions of bad data, or the two LMC globular 
clusters that fall in the survey footprint (Reticulum and ESO121-SC03). Figure \ref{f:mlcirc} (upper panel) shows the one- and 
two-dimensional marginalized posterior probability distributions for our model parameters, and Table \ref{t:ml} lists the most-likely 
values for these quantities. For consistency with the literature, in which the position angle of the line of nodes is usually taken 
counterclockwise from the north, we follow \citet{vdm:01a} and report our results in terms of $\Theta=\theta-90\degr$. The formal 
uncertainties for all parameters are very small, showing that the algorithm has converged well; however it is likely that the true 
uncertainties, including systematics, are substantially larger.

Our inferred values for the inclination and scale length are consistent with estimates from previous work on the outer LMC 
disk, although some degree of variation is seen between measurements in the literature due to the different regions covered by 
various studies, and the use of different tracer populations.  For example, \citet{vdm:01a} and \citet{vdm:01b} find $i = 34.7\degr \pm 6.2\degr$
and $r_s \approx 1.3-1.5$ kpc by tracing AGB and RGB stars over the radial range $\sim2\degr-7\degr$, while \citet{weinberg:01} find 
$i = 24.0\degr \pm 0.3\degr$
and $r_s = 1.42 \pm 0.01$ kpc over a similar radial range using a variety of old and intermediate age tracers. \citet{saha:10} 
measure $r_s = 1.15$ kpc by tracing old main sequence turn-off stars in a stripe spanning $\sim6\degr-16\degr$ due north of the
LMC and assuming the geometry of \citet{vdm:01a}. More recently, \citet{balbinot:15} obtain $i = 32.9\degr\pm0.4\degr$
and $r_s = 1.41 \pm 0.01$ kpc by tracing old main sequence turn-off stars in a region spanning a similar radial range to the north-east. 

\begin{figure}
\begin{center}
\includegraphics[width=\columnwidth]{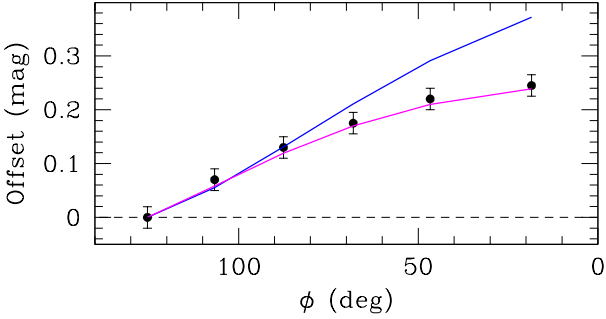}\\
\end{center}
\caption{Relative distance measurements as a function of position angle $\phi$ for our six outer disk segments, along with the
profiles for our inclined circular and elliptical disk models (blue and magenta lines, respectively). The black points represent the mean
value of the main sequence turn-off and red clump distance measurements for each segment.
\label{f:dcomp}}
\end{figure}

The lower panel of Figure \ref{f:mlcirc} shows our model plotted on the LMC density map.
There is good general correspondence between the shape of the model and the observed isodensity contours. Our inferred value for
$\Theta$ implies the line of nodes lies $\sim 20\degr$ from vertical on the map.
In Figure \ref{f:dcomp} we plot the line-of-sight distance profile for the best-fit model (blue line). Although the model does exhibit a clear
distance gradient from east to west, the gradient is too steep in the sense that the magnitude offset between the far east and far 
west of the survey area appears too large. 
 
There are two reasons for this discrepancy. First, the position angle of the line of nodes is not a good match for the observed distance
profile. The sinusoid plotted in Figure \ref{f:dgrad} implies that the line of nodes should sit at an angle $\sim 5\degr$ east of
north; however that for our inclined circular disk model is oriented $25\degr$ further westwards. Forcing the line of nodes in
the model to match that implied by the distance profile would move the blue curve left and downwards on the plot (recall that 
all our distance measurements are relative to the easternmost segment).  However, this would then mean that the shape of the
model would not fit the isodensity contours so well -- for an inclined circular disk, the orientation of the line of nodes must coincide 
with the direction of the major axis of the (elliptical) isopleths on the sky.
  
Second, regardless of the orientation of the line of nodes, the extremes of the line-of-sight distance in our model (at any azimuth, not 
just within the survey footprint) over angular separations from the galaxy centre commensurate with those for our six segments, span a 
range of $\approx 0.45$ mag in distance modulus, which is $\sim 30$ per cent larger than the peak-to-peak amplitude suggested by 
our observed distance profile. This implies that the best-fit inclination is too high; however, a reduction in the inclination would again mean 
that the shape of the model on the sky would not fit the isodensity contours so well, as this parameter controls the degree of ellipticity
on the sky.
 
Retaining the assumption of a planar geometry, one way of resolving these issues is if the outer LMC disk is intrinsically 
elliptical rather than circular, perhaps due to tidal stresses acting on the disk \citep[see e.g.,][]{vdm:01b}.
To test the feasibility of this hypothesis we replaced the circular disk in 
our model with an ellipse defined on the $(x,\,y)$ plane by
\begin{equation}
\frac{(x \cos\psi-y \sin\psi)^2}{a^2}+\frac{(x \sin\psi+y \cos\psi)^2}{a^2(1-e)^2}=1\,.
\end{equation}
The ellipse has semi-major axis $a$ and ellipticity $e=(1-q)$ where $q$ is the axis ratio, and is rotated on the plane by
an angle $\psi$ which, for consistency with $\phi$ and $\theta$, we measure counterclockwise from the ``west'' (i.e., 
when $\theta$ and $i$ are zero, positive $\psi$ gives rotation north of west on the sky). The scale length is defined along the 
major axis of the ellipse on the disk plane. The extra parameters in this model mean that the orientation of the density contours 
on the sky is decoupled from the position angle of the line of nodes; moreover, the apparent ellipticity of the isopleths is at least 
partly controlled by the intrinsic shape of the elliptical disk rather than solely by the inclination. 

We tried to fit the full six-parameter elliptical model to our data; however because the survey footprint covers less than
$40$ per cent of the outer LMC disk, several of the parameters are largely degenerate and the algorithm fails to converge. 
We thus attempt to simplify the problem by fixing the orientation of the line of nodes to the value $\Theta=185\degr$ implied
by our distance profile -- essentially we force the phase of the sinusoidal distance variation to agree with the observations
and then search for a solution that matches the shape of the density isophotes on the sky.

\begin{figure}
\begin{center}
\includegraphics[width=\columnwidth]{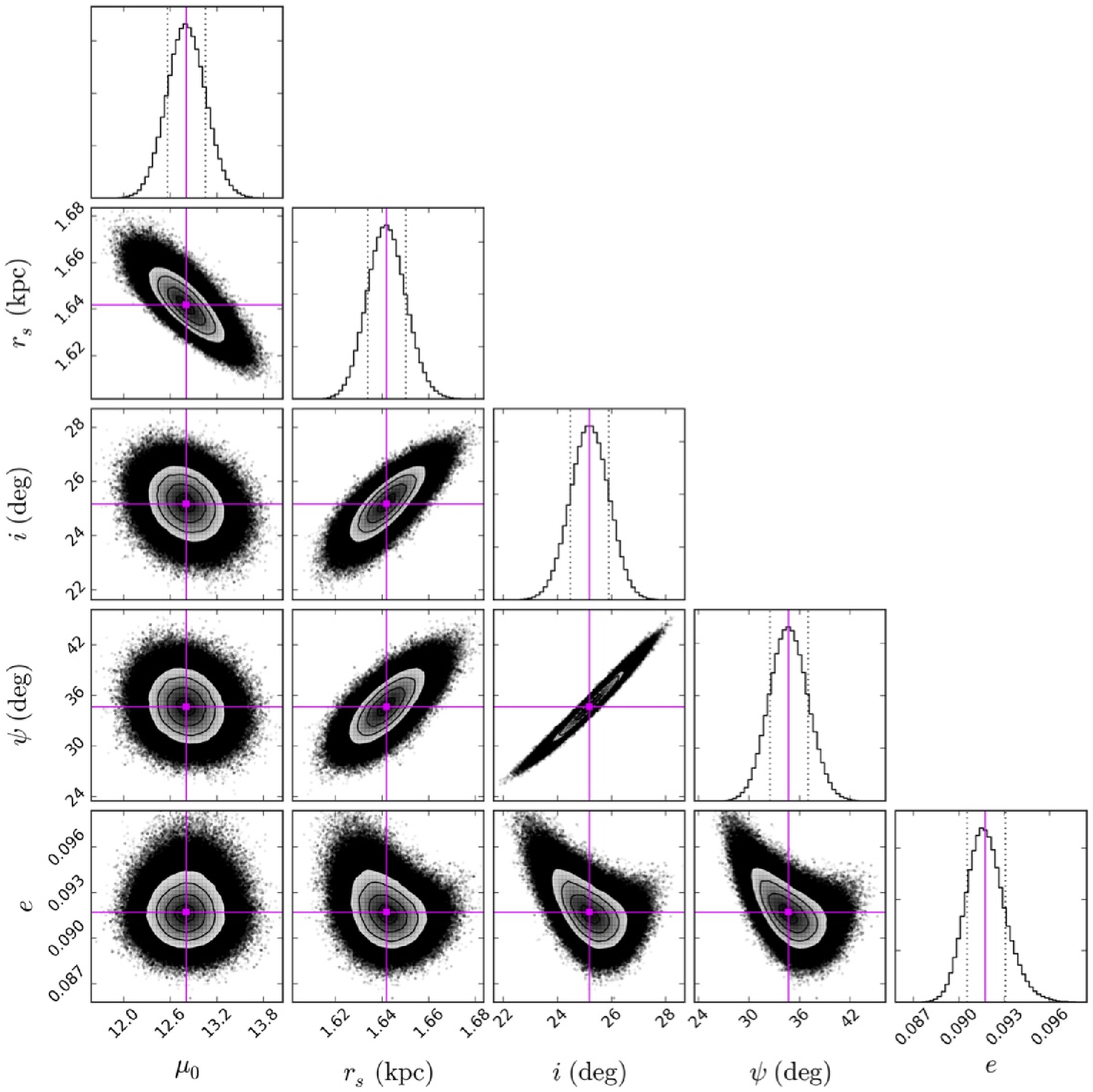}\\
\vspace{1mm}
\includegraphics[width=\columnwidth]{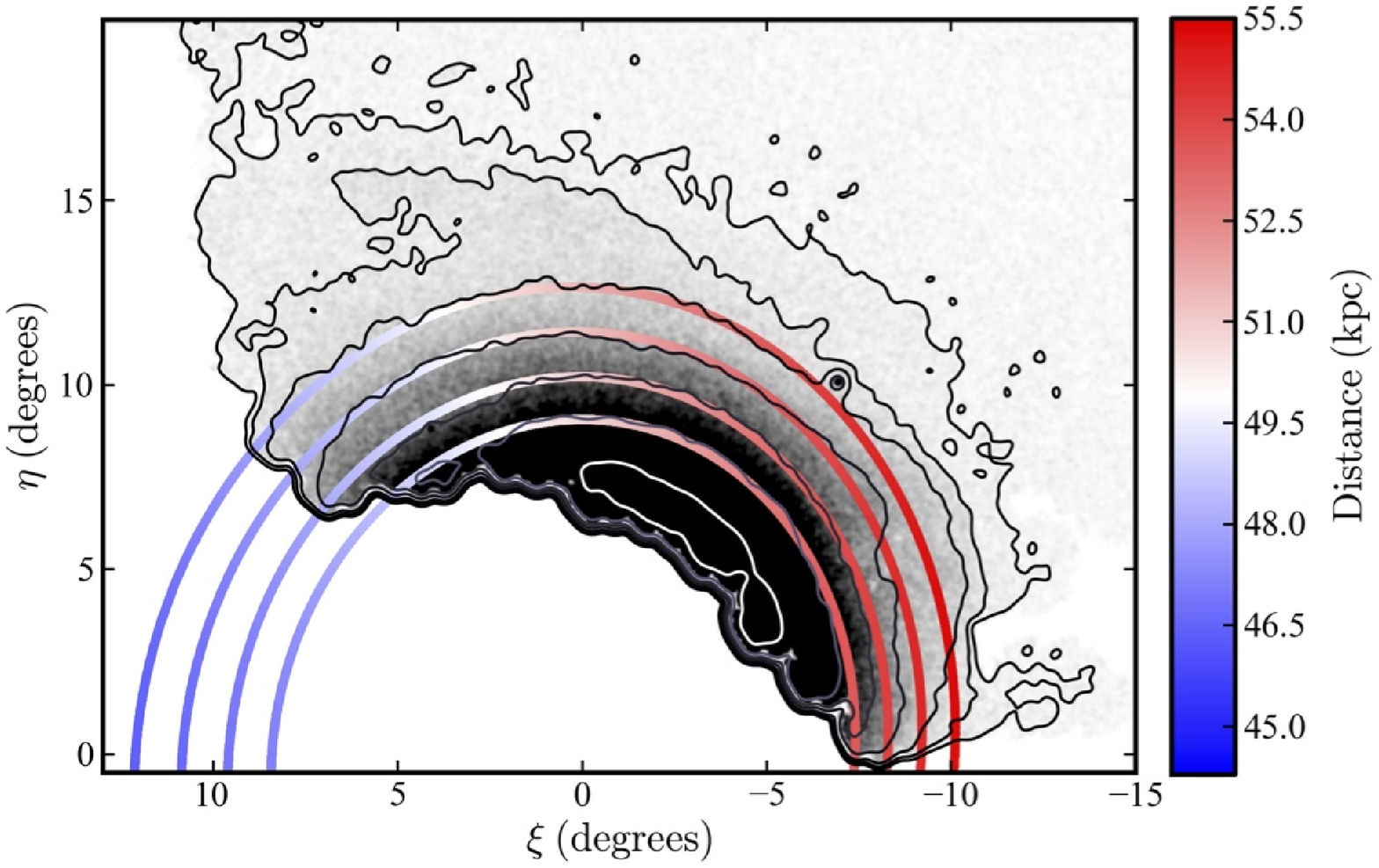}
\end{center}
\caption{Same as Figure \ref{f:mlcirc} except for our inclined elliptical exponential disk model. The five parameters are the central
density $\mu_0$, the scale length along the major axis $r_s$, the inclination angle $i$, the position angle of the elliptical disk
on the LMC plane $\psi$, and the ellipticity of the disk $e$. In the lower panel the model is plotted at radii of $8.5$, $9.5$,
$10.5$, and $11.5$ kpc in the plane of the disk. Again, there is good general correspodence between the shape of the model and 
the isodensity contours. As previously, north is up and east to the left. 
\label{f:mlell}}
\end{figure}
 
The upper panel of Figure \ref{f:mlell} shows the marginalized posterior distributions 
for this five-parameter problem; the most-likely values of the parameters are listed in Table \ref{t:ml}.  The scale length is 
similar to that for our circular model.  The inclination is substantially lower, although it is still well within the range of inclination 
measurements found in the literature at smaller radii \citep[see e.g.,][and the compilation therein]{subra:10} and at larger
radii as discussed above. Interestingly, our value for $i$ matches quite closely with a linear extrapolation of the radial variation 
in this quantity seen out to $\sim 7\degr$ by \citet{vdm:01a} (their Figure 8). Finally, the intrinsic ellipticity of 
our model disk is somewhat lower than that of \citet{vdm:01b}, which has $e \approx 0.31$ out to $\sim 7\degr$; indeed our 
model is not far from being circular.
 
The bottom panel of Figure \ref{f:mlell} shows our elliptical disk model plotted on the LMC density map. As with our circular 
model, there is in general very good agreement between the shape of the model and the observed isodensity contours.
The line-of-sight distance profile for the elliptical model is plotted with a magenta line in Figure \ref{f:dcomp}. This is an
excellent match to the observed distances. In particular, the extremes of the line-of-sight distance in the model at galactocentric
radii comparable to those for our six segments span a range of $\approx 0.34$ mag in distance modulus, which is entirely
consistent with the peak-to-peak amplitude inferred observationally. Although we fixed the position of the line
of nodes to match the observations, it is encouraging that the correct amplitude of variation arises naturally from 
the maximum likelihood algorithm even though this is simply trying to match the observed stellar density on the sky without
incorporating any explicit distance dependence.

In summary, while an inclined circular disk model can describe the observed density distribution of the LMC outskirts on the
sky, it cannot simultaneously reproduce the observed azimuthal line-of-sight distance profile.  To do this successfully
requires a more general inclined elliptical disk model.  Even so, there is still one prominent discrepancy between this model
and our observations.  Equation \ref{e:dist} suggests that the line-of-sight distance to points on the line of nodes
should be $D = D_0/\cos\rho \approx 18.52$, whereas we measure a distance modulus of $\sim 18.35$ in the relevant azimuthal 
segment. While there are certainly systematic effects to be concerned about in the derivation of this distance, these would have 
to be quite substantial ($\approx 0.15-0.2$ mag) in order to completely account for this offset. One possibility is that 
the LMC geometry becomes non-planar at large galactocentric separations, and indeed a number of previous works have 
seen evidence for such deviations \citep[e.g.][]{nikolaev:04,subra:10,balbinot:15}.
Flaring and/or warping of the outer LMC disk could explain the difference between our absolute distance scale and that
predicted by the elliptical disk model. 

Twisting of the line of nodes with galactocentric separation is a key signature expected if the LMC disk is warped 
\citep[e.g.,][]{nikolaev:04}. Most studies of the inner LMC infer a position angle for the line of nodes to be in the range 
$\sim140\degr-170\degr$ east of north \citep[see e.g., the compilation in][]{subra:10}, although \citet{vdm:01a} found 
a value of $122.5\degr \pm 8.3\degr$ east of north out to $\sim 7\degr$. Our distance profile suggests 
$\Theta \approx 185\degr$ at large galactocentric radii, and is thus perhaps indicative of just such a twist. 
 
\begin{figure*}
\begin{center}
\includegraphics[width=85mm]{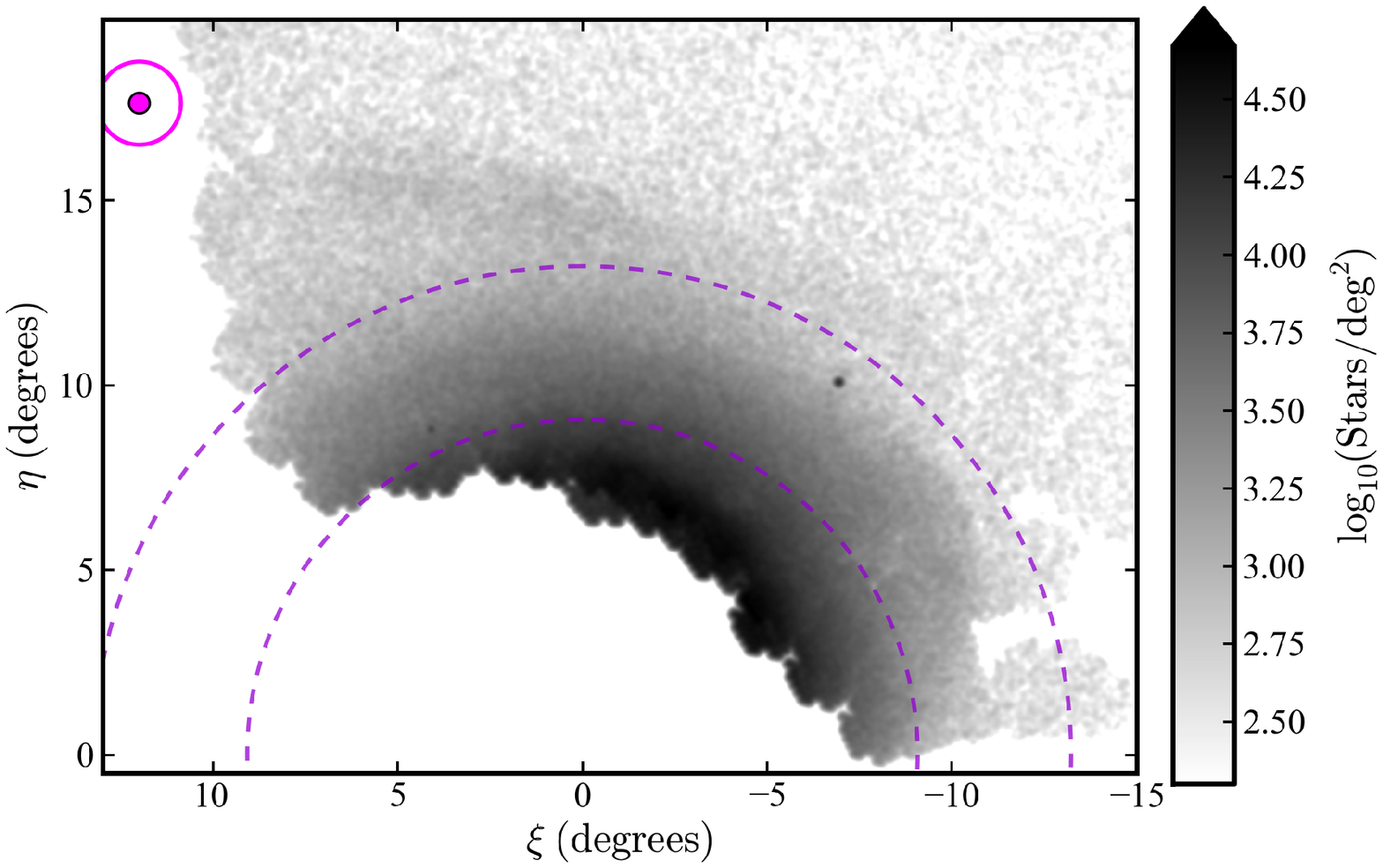}
\hspace{1mm}
\includegraphics[width=85mm]{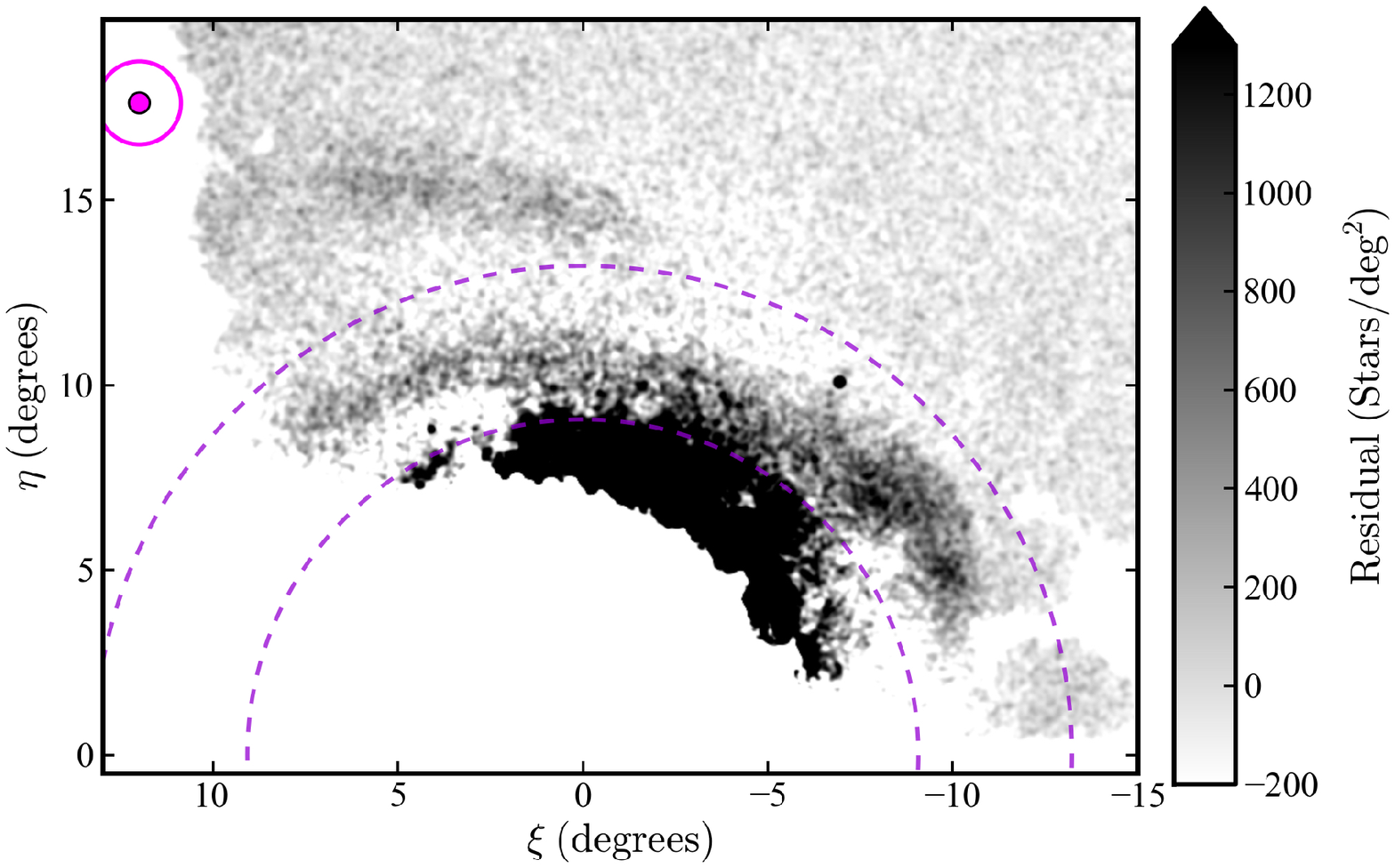}
\end{center}
\caption{Spatial density maps of the outer LMC disk. {\bf Left panel:} A map similar to that shown in Figure \ref{f:map} but with a logarithmic scaling to reveal features in the region interior to $\sim 13\degr$. The LMC globular clusters Reticulum and ESO121-SC03 are clearly visible, at $(\xi,\,\eta)\approx(-7,\,10)$, and $\approx(4,\,8)$, respectively.  Also apparent are several possible deviations from a smooth distribution, primarily in the form of sharp edges near $(\xi,\,\eta)\approx(-10,\,7)$ and $\approx(6,\,10)$. The two dashed circles indicate radii of $9\degr$ and $13\degr$ from the LMC centre, corresponding to the range over which our planar disk models were fit. {\bf Right panel:} A map showing, with a linear scale, residual star counts once our best fitting elliptical disk model has been subtracted, together with a uniform background (see Section \ref{ss:profile}, below). In addition to our remote stream-like feature, widespread deviations from a smooth planar disk are clearly evident between $9\degr-13\degr$.  Within $9\degr$ the model does not provide a good description of the data due to the presence of additional young stellar populations in our CMD selection box.
\label{f:diskmaps}}
\end{figure*}
 
It is also worth noting that the outermost density contours in the lower panels of Figures \ref{f:mlcirc} and \ref{f:mlell}
deviate noticeably from a purely elliptical shape. This does not just occur in the region of the stream-like overdensity -- for example, 
there is a clear outwards bulge in the contours to the north-west, near $(\xi,\,\eta) = (-10,\,7)$. In the left panel of Figure \ref{f:diskmaps} 
we show a logarithmic stellar density map of the outer disk, revealing a sharp edge in this region, as well as additional edges at other 
azimuthal angles and radii -- for example, near $(\xi,\,\eta) = (6,\,10)$. These features suggest that not only is the outer LMC disk 
possibly distorted, its structure may not be smooth.
 
To explore this idea in more detail, we construct in the right hand panel of Figure \ref{f:diskmaps} a map showing the residual
differences between the observed stellar density and our best-fitting smooth elliptical disk model. In the region over which the 
best-fitting model was calculated (i.e., at radii between $9\degr-13\degr$ from the LMC centre), substantial positive residuals are 
evident. These take the form of a low suface brightness ring-like structure spanning all azimuthal angles in the survey footprint, but 
densest to the west.  The sharp-edged features seen in the logarithmically scaled map of the disk are due to this structure. Its nature is 
not clear from the presently-available data -- the structure could represent an overdensity in the disk itself, akin to a spiral arm, or, 
more speculatively, it may represent a stellar stream of some description, perhaps due to an accreted satellite\footnote{It is interesting 
that the peculiar remote LMC globular clusters Reticulum and ESO121-SC03 \citep[e.g.,][]{mackey:04,mackey:06} sit on either edge of the
structure.}. Further investigation is clearly warranted.

Inside $9\degr$ our model does not provide a good fit to the data.  This is largely due to the presence of young stellar populations
at such radii, as discussed in Section \ref{ss:cmds}. Outside $13\degr$ the model does an excellent job of describing the structure
of the LMC disk, at least to the east of $\xi=0$, even though data at such large radii were not used to calculate the model. To the west
of $\xi=0$ our new stream-like feature is clearly visible stretching towards the Carina dwarf. The map clearly shows that this
feature ``terminates'' due north of the LMC-- further east and further south the data are well described by our planar disk model.

\subsection{Outer surface density profile and the extent of the LMC}
\label{ss:profile}
\begin{figure}
\begin{center}
\includegraphics[width=\columnwidth]{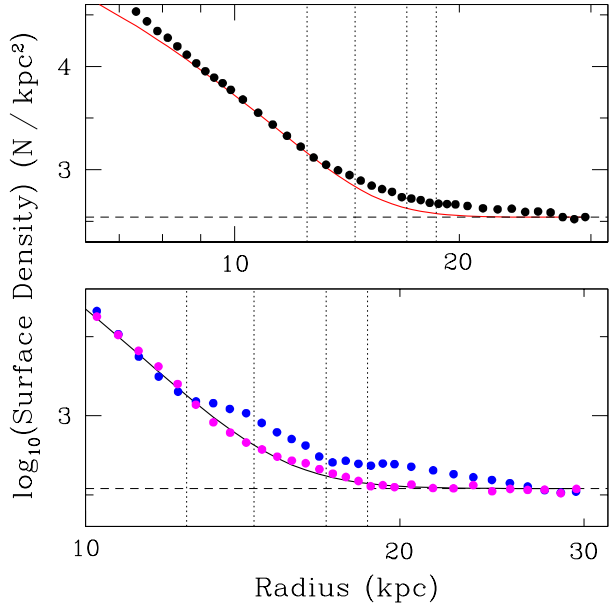}
\end{center}
\caption{Radial surface density profiles for stars falling in our main sequence turn-off selection box on the CMD. {\bf Upper panel:} 
Density profile for all stars. {\bf Lower panel:} Density profiles for stars lying to the east and west of $\xi = 0$ (blue and magenta points,
respectively). In both panels the prediction of our best-fitting elliptical disk model, plus the background, is marked (red line in the 
upper panel; black line in the lower panel). Also marked in both panels is the background level (black dashed line) as determined 
empirically using the western profile, and the four key radii on the LMC disk plane discussed in the text (vertical dotted lines). 
These are: (i) $12.5$\ kpc ($\approx 13.5\degr$), which marks the innermost edge of the main stream-like overdensity and the 
radius at which the eastern and western profiles begin to diverge; (ii) $14.5$\ kpc ($\approx 15.7\degr$), which marks the outer 
edge of the faint features in the region to the west of $\xi=0$; (iii) $16.5$\ kpc ($\approx 17.8\degr$), which marks the outer edge 
of the main stream-like overdensity in the region to the east of $\xi = 0$; and (iv) $18.5$\ kpc ($\approx 20\degr$), which marks 
the radius beyond which we do not see evidence for LMC populations in any direction.
\label{f:profile}}
\end{figure}

In the upper panel of Figure \ref{f:profile} we plot the radial surface density profile for stars in our main sequence turn-off CMD 
selection box, where the radius is measured along the major axis in the plane of the elliptical disk from the best-fitting model 
described above.  We corrected all star counts using the empirical relation in Equation \ref{e:corr}.  The predictions of the model 
(including a background contribution, see below) are marked with a red line. This provides an excellent fit over the radial range we 
used for the input data ($\sim 9\degr-13\degr$). This is despite the presence of the low surface brightness structure identified
in Figure \ref{f:diskmaps}, which is disguised by the azimuthal averaging inherent in the construction of Figure \ref{f:profile},
plus the logarithmic scale.

As previously noted, for radii inside $\sim 8.5$\ kpc ($\approx 9\degr$) the density is higher than the model, due to increasing 
contamination from younger LMC populations. Outside $\approx 12.5$\ kpc ($\approx 13.5\degr$), the density again becomes noticeably
higher than predicted by the model, due to the presence of the stream-like overdensity and contamination of the CMD selection
box (primarily by foreground sources). At very large radii, the density profile asymptotes 
to an approximately constant background level (marked with a horizontal dashed line); we discuss this in more detail below.

The lower panel of Figure \ref{f:profile} shows the radial surface density profile split into two -- blue for stars sitting to the east
of a line running due north from the LMC centre (i.e., all stars east of $\xi = 0$), and magenta for stars sitting to the west of this line.
Although the profiles match quite well at inner radii, they deviate substantially outside $\approx 12.5$\ kpc (i.e., $\approx 13.5\degr$). 
We mark this radius with a vertical dotted line in the plot. This deviation is the signature of the main body of our new stream-like
overdensity, which lies almost entirely to the east of $\xi = 0$. Outside $12.5$\ kpc, the density profiles sit apart for almost the 
full extent of our measurements.

\begin{figure}
\begin{center}
\includegraphics[width=\columnwidth]{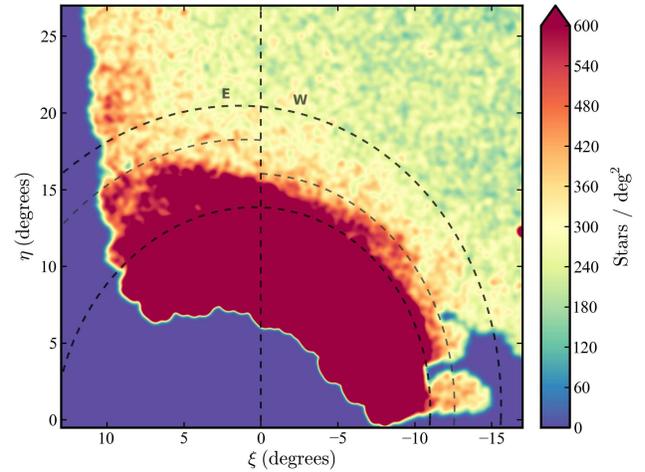}
\end{center}
\caption{A final incarnation of our stellar density map, with the colour-scale set to saturate at $600$ stars deg$^{-2}$
and a large degree of smoothing (using a Gaussian kernel of width $\sigma = 10\arcmin$). As before, north is up and east 
to the left. The dashed ellipses correspond to the four key radii plotted in Figure \ref{f:profile} and described in the text.
The innermost ellipse marks a radius in the disk plane of $12.5$\ kpc, while the outermost ellipse marks $18.5$\ kpc. 
Between these inner and outer ellipses, a radius of $14.5$\ kpc is marked in grey to the west of $\xi=0$, and a radius of 
$16.5$\ kpc is marked in grey to the east. For convenience the line $\xi=0$ is also plotted.
\label{f:faint}}
\end{figure}
 
We interpret these profiles with the aid of our stellar density map, which is presented in Figure \ref{f:faint} with a high degree of 
smoothing and a colour scaling that emphasises very low-density features at large radii. The innermost dashed ellipse on this map 
corresponds to a radius of $12.5$\ kpc; it is easy to see why the profiles deviate so strongly immediately outside this bound. 
The western profile (magenta) decreases smoothly out to a radius of $\sim 18.5$\ kpc, where it flattens to a constant background. 
This radius is marked with the outermost dashed ellipse on the density map, and the outermost vertical dotted line on the profile. 
The density map shows that to the west of $\xi = 0$ and between $12.5$\ kpc and $18.5$\ kpc, there are clearly still LMC stars 
present. Most of these remote stars sit within $\sim 14.5$\ kpc (also marked), which traces the approximate outer edge of the 
faint western structure that we remarked upon in Section \ref{ss:map}; however, there are also even fainter features 
present in the range $14.5-18.5$\ kpc. These extremely low surface density populations are likely to be associated with the 
LMC as they broadly follow the same elliptical shape on the sky. As noted in our discussion of Figure \ref{f:diskmaps} above,
in this region to the west of $\xi = 0$ our exponential disk model, which was derived using stars between $9\degr-13\degr$, 
provides quite a good fit to the density profile over the full radial range out to $18.5$\ kpc ($\approx 20\degr$) despite the apparent 
presence of the faint western structure.
 
\begin{figure}
\begin{center}
\includegraphics[height=75mm]{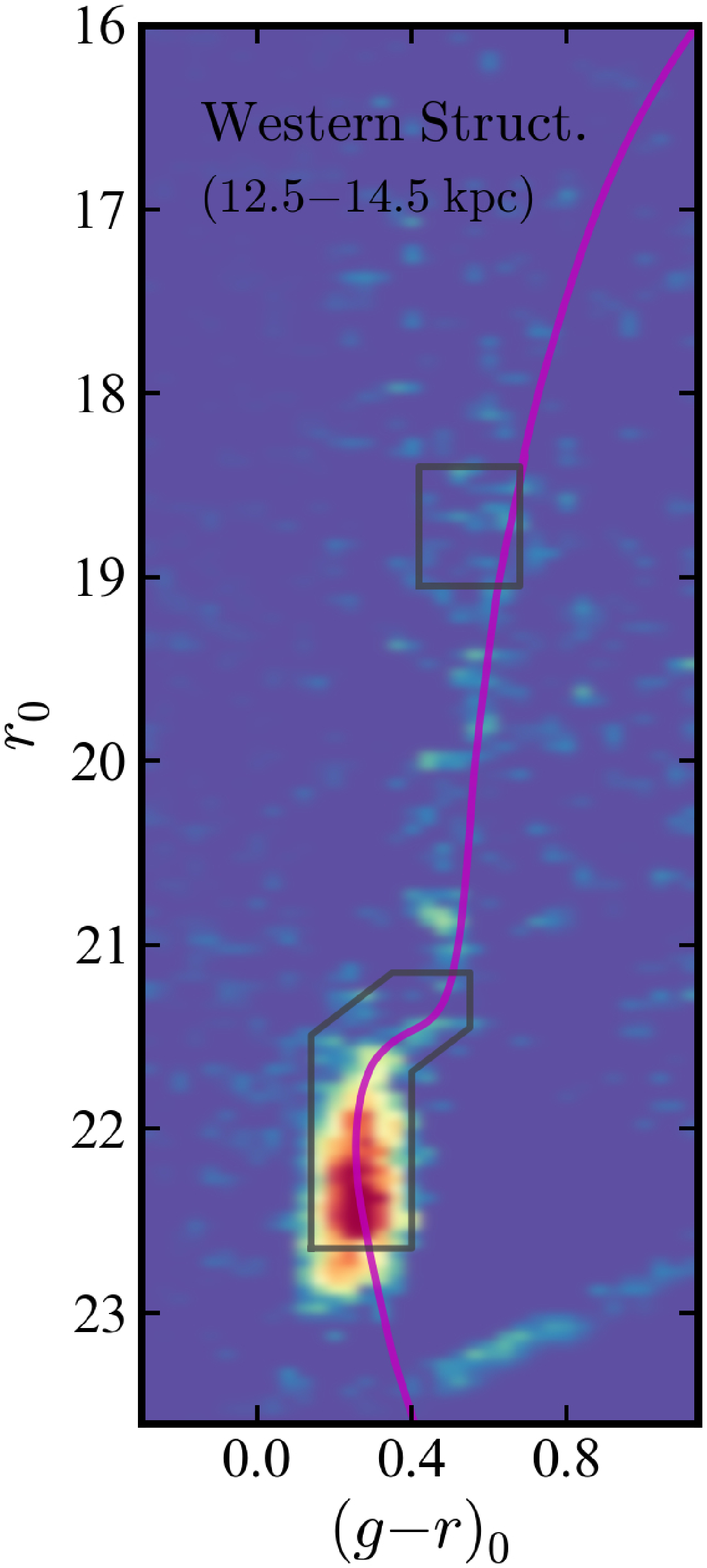}
\hspace{-2mm}
\includegraphics[height=75mm]{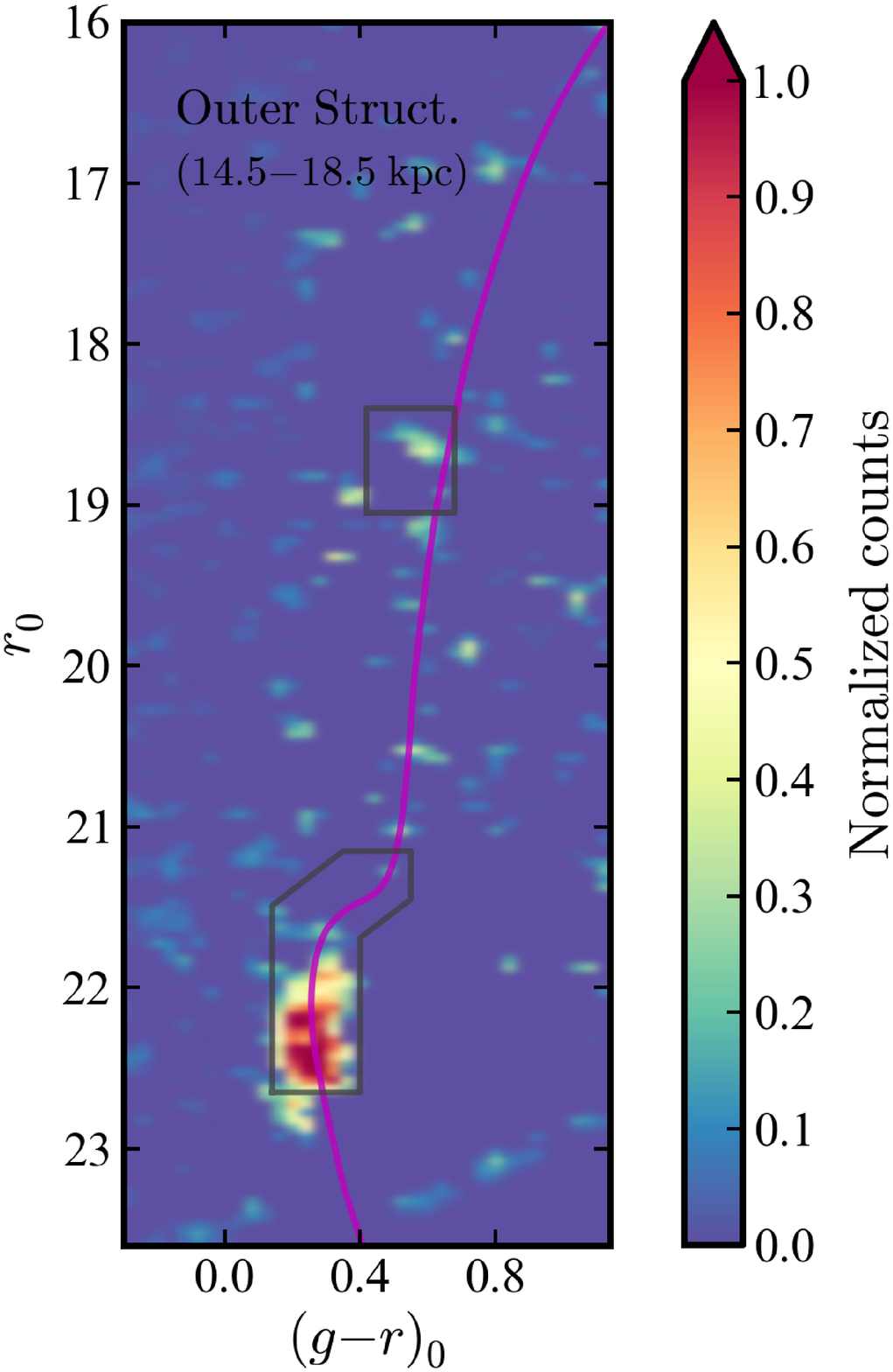}
\end{center}
\caption{Field subtracted Hess diagrams for stars lying to the west of $\xi = 0$ and between $12.5-14.5$\ kpc ({\bf left}) and 
$14.5-18.5$\ kpc ({\bf right}). The fiducial isochrone and CMD selection boxes are the same as in Figures \ref{f:cmd} and \ref{f:streamcmd}.
LMC stellar populations are evident even to very large radii.
\label{f:outercmd}}
\end{figure}

To investigate the region to the west of $\xi = 0$ in more detail, we constructed field subtracted Hess diagrams in the radial range 
$12.5-14.5$\ kpc and $14.5-18.5$\ kpc.  We utilised exactly the same procedure as that which we
previously used to construct the field subtracted Hess diagrams for the main body of the stream-like overdensity 
(Figure \ref{f:streamcmd}), but in this case adopting field boxes of appropriate area from the western region just outside $18.5$\ kpc. 
These Hess diagrams are presented in Figure \ref{f:outercmd}, and reveal that LMC-like populations are clearly visible even to very large 
radii as inferred from the stellar density map and radial profile. To the precision possible with such low-density CMDs, the populations
in this region appear to match both the LMC disk within $12.5$\ kpc ($\approx 13.5\degr$) and the main stream-like overdensity.  

Returning to the stellar density map and now considering the region to the east of $\xi = 0$, the picture is complicated by the
presence of foreground contamination that increases noticeably towards the edge of the survey footprint. This is due to the 
Milky Way disk, which lies roughly parallel to the edge of the survey footprint another $\sim 20\degr$ further east.
We have already discussed the issue of foreground contamination in Section \ref{ss:effects}; here, we note that these
contaminants make a non-negligible contribution to the surface density profile outside $\approx 16.5$\ kpc (marked with a 
grey dashed ellipse on the map and a vertical dotted line on the profile), which is the radius within which the main stream-like
features are concentrated east of $\xi=0$.  The blue profile in Figure \ref{f:profile} should not be trusted outside $16.5$\ kpc; 
this contamination is the reason why the profile for all stars, in the upper panel of the plot, does not converge to the background 
level (determined using the profile to the west of $\xi=0$) until very large radii.

In summary, we find that to the west of $\xi=0$ the LMC radial surface density profile shows clear evidence for very diffuse populations
reaching out to $18.5$\ kpc ($\approx 20\degr$). These may also be present to the east of $\xi=0$, but we cannot be certain due to increasing 
contamination from the Galactic foreground. Our observation of such remote LMC populations is consistent with the spectroscopic detection 
in many directions of probable LMC giants out to radii of $\approx 20\degr$ by \citet{majewski:00,majewski:09} and \citet{munoz:06}, and the
recent blue-horizontal branch star maps of \citet{belokurov:16}. Our {\it overall} LMC surface density profile shows a gentle break outside 
a radius of $12.5$\ kpc from the best-fit exponential fall-off measured in Section \ref{ss:geom} to a flatter decline.  This is quite
similar to the profile observed by \citet{majewski:09} -- although they assume an exponential scale length of $1.6\degr \approx 1.5$\ kpc 
and a break radius of $9\degr \approx 8.5$\ kpc, their spectroscopic data points (top panel of their Figure 2) exhibit a substantial spread 
and would also be compatible with our slightly longer scale length of $1.64$\ kpc and larger break radius of $12.5$\ kpc\ $\approx 13.5\degr$. 
In addition we find that 
these remote populations have characteristics similar to those of stars in the outer LMC disk and in the stream-like overdensity we have 
discovered stretching to the east. This observation is again compatible with the spectroscopic observations by \citet{majewski:09}, which 
show an apparently flat metallicity profile out to $\sim 20\degr$, with a mean value of $[$Fe$/$H$] \sim -1.0$.  

While \citet{majewski:09} interpret the break in their profile to indicate the presence of a classical halo around the LMC, the break 
in our overall profile can be fully explained by the stream-like feature extending east -- when we construct a profile using only stars to
the west of $\xi=0$, where this overdensity is not present, it is well described by our best-fit exponential disk model over the full radial 
range out to $18.5$\ kpc ($\approx 20\degr$).  This does not rule out the presence of a stellar halo -- it is important to note that
the fields observed by \citet{majewski:09} span a signfiicantly wider range in azimuth than does the present DES footprint, and, moreover, 
the recent maps by \citet{belokurov:16} tracing blue horizontal branch stars suggest an LMC halo component almost certainly exists. 
However, it does indicate that (i) the LMC exponential disk may extend further than previously suspected, at least in some directions, 
and/or (ii) the transition to a halo-like population is rather smooth such that it does not lead to a noticeable break in the profile
(at least to the north-west). If the disk indeed extends out to such large radii, this would be in contrast to some previous estimates
for its truncation radius, which lie in the range $\sim 13-15$\ kpc \citep[e.g.,][]{saha:10,balbinot:15}. We have already noted in 
Section \ref{ss:map} how the sharp drop off in density $\approx 15\degr-16\degr$ due north of the LMC could explain the apparent 
trunction in the profile of \citet{saha:10}, and we suspect a similar effect could explain the ``edge'' seen at $\sim 13.5\pm 0.8$\ kpc 
to the north-west by \citet{balbinot:15}. In this direction we observe the faint western structure extending to $\sim 14.5$\ kpc; it may 
be that our CMD selection box, which traces much more of the main sequence and main sequence turn-off region than that of 
\citet{balbinot:15}, has facilitated the detection of extremely diffuse populations outside this radius -- note that on the CMD in 
Figure \ref{f:outercmd}, LMC stars beyond $14.5$\ kpc are only clearly detected below the turn-off. 

\section{Discussion \& Summary}
\label{s:discussion}
We have used year one DES imaging to study the structure of the northern periphery of the LMC. We find evidence for a substantial
stream-like stellar overdensity stretching from $\approx 13.5\degr$ due north of the LMC centre towards the east. Although this is, 
broadly speaking, a similar direction to that of the leading arm of the Magellanic H{\sc i} stream, there is no obvious correspondence 
between the two -- the gaseous feature is {\it much} wider and more extended \citep[see e.g.,][]{nidever:08,nidever:10}, and there is at 
present no suggestion that our substructure comprises a stellar component to it. 

The width of our overdensity is $\sim 2\degr$, or $\approx 1.5$\ kpc, and it spans an arc on the sky that is, at minimum, $12.5\degr$ 
long -- corresponding to a physical length greater than $\sim 10$\ kpc.  This assumes that the feature passes in front of the Carina 
dwarf galaxy around $20\degr$ north-east of the LMC, possibly explaining why a number of previous studies of this object have found 
evidence for LMC populations in their field of view \citep[e.g.,][]{mcmonigal:14} that exhibit cold kinematics \citep[e.g.,][]{munoz:06}.
We can think of three possible scenarios for the origin of this structure:
\begin{itemize}
\item{It might be the remains of a disrupted satellite of the LMC. Recent searches have uncovered a substantial 
number of low-luminosity dwarf galaxies in the vicinity of the LMC \citep{koposov:15,bechtol:15,kim:15a,kim:15b,martin:15}, some
of which may well be bound to the Magellanic Clouds \citep[e.g.,][]{deason:15}.}
\item{It may be an overdensity in the remote outskirts of the LMC disk itself -- perhaps akin to an extended spiral arm. The recent
diffuse light imaging by \citet{besla:16} has revealed similar structures extending to $\approx 8\degr$ from the LMC centre.}
\item{It could be comprised of material that is being stripped from the outer LMC disk due to the tidal force of the
Milky Way, or due to a recent close encounter with the SMC.}
\end{itemize}

The observed properties of the substructure help shed more light on these possibilities. We find that the integrated luminosity of its brightest
portion is $M_V \approx -7.4$.  This sits in between the Hercules dwarf ($M_V \sim -6.6$) and the Draco and Ursa Minor
dwarfs ($M_V\sim -8.8$) in the Milky Way system \citep[e.g.,][]{mcconnachie:12}, although we note that our measurement is a lower limit 
since we have not traced the full extent of the overdensity on the sky. Our field-subtracted Hess diagrams show that the stellar populations in 
the substructure are predominantly old, as would be expected for a faint dwarf, but, critically, they also have $[$Fe$/$H$] \sim -1$.  
This is much more metal-rich than would be expected for such a low luminosity system according to the luminosity-metallicity relation 
\citep[e.g.,][]{kirby:11} -- Draco, Ursa Minor, and Hercules all have $[$Fe$/$H$] \la -2$. It could be that the putative satellite was initially 
much more massive and has suffered extensive tidal destruction; however the original system would need to have been unfeasibly large -- 
nearly two orders of magnitude more luminous than at present, comparable to the Fornax dwarf with $M_V \approx -13.4$.

Our photometry reveals that, in fact, the stellar populations seen in the stream-like feature are an excellent match 
for those found in the outer LMC disk. This is strongly suggestive that it is comprised of disk material -- perhaps stars being tidally
stripped from the disk, or an overdensity in the structure of the disk itself. Many previous studies have found non-axisymmetric features
in density maps of the LMC -- there are spur-like structures, possibly spiral arms, in the inner regions \citep[e.g.,][]{zaritsky:04}; while
at intermediate radii there are spur-like and shell-like features present \citep[e.g.,][]{irwin:91}, some of which sit towards the north 
and progress from west to east \citep[e.g.,][]{devau:72}. More recently, \citet{besla:16} have revealed arc-like structures in the disk extending
to $\sim 8\degr$. Furthermore, we clearly observe additional structure of some description in the LMC 
disk at radii of $9\degr-12\degr$ (see Figure \ref{f:diskmaps}), although the nature of these features is not clear at present. Despite the 
fact that we detect no young populations at the radius of our overdensity, such that it is probably not a spiral arm in the traditional sense, 
it is not difficult to imagine a scenario in which one or more close encounters between the LMC and SMC \citep[as in][]{besla:16}, or between 
the LMC and the Milky Way, could give rise to transient ``ripples'' or perturbations in the outer LMC disk, or indeed remove material entirely 
into a stellar stream.

In either of these cases, it might be expected that the outer LMC disk ought to show some other signs of distortion.  We have 
observed tentative evidence for this. By constructing models of inclined planar disks we find that the observed stellar density of the 
outer LMC disk on the sky, and its azimuthal distance profile, can only be simultaneously explained by an elongated disk. The necessary 
ellipticity is only mild, but might indicate the presence of tidal distortion \citep[see the discussion in][]{vdm:01b}. It is also relevant 
that the orientation of the elliptical isodensity contours of the outer LMC is such that the major axis is aligned very close to the 
north-south direction on the sky, coinciding with the apparent location of the origin of the stream due north of the LMC centre. 

In addition to this, the position angle of the line of nodes in our best-fitting elliptical disk model is twisted relative to all previous 
determinations from more central regions, which may indicate a warp in the disk. Our measured inclination is also lower than most 
estimates derived from the inner LMC; indeed, it follows the general trend of decreasing inclination with radius observed by 
\citet{vdm:01a}. Finally, there is an offset in our absolute distance measurement to the outer LMC disk, which we find to be about 
$0.15-0.20$ mag brighter than predicted by our best-fitting planar model. While much of this offset can likely be attributed to 
systematics associated with deriving distances using isochrone models, it might also reflect warping or flaring of the disk, or some 
other kind of distortion along the line of sight. This type of non-planar behaviour at large radial distances is again possibly 
indicative of tidal perturbations or distortions in the outer disk.

If our substructure is indeed comprised of stripped disk material, then its origin might be due to the tidal force of the Milky Way. 
This is the argument put forward by \citet{vdm:01b} to explain the (substantial) intrinsic ellipticity they infer for the LMC disk --
the orientation that they measure for the elongation points within $\sim 6\degr$ of the direction of the Galactic centre, which 
sits at a position angle of $183.7\degr$ relative to the LMC centre. Our results are in excellent agreement with this finding -- the 
position angle of the line of nodes in our best elliptical disk model sits at $185\degr$, while the ellipticity on the sky is almost
perfectly aligned in the north-south direction (see e.g., the lower panel of Figure \ref{f:mlell}). Moreover, as expected in the case 
where our substructure is a stream induced by the tidal force of the Milky Way, its direction is roughly perpendicular to the 
disk elongation and quite well aligned with the latest {\em Hubble Space Telescope} proper motions, which are 
$\mu_W = -1.910\pm 0.020\,$mas$\,$yr$^{-1}$ and $\mu_N = 0.229\pm 0.047\,$mas$\,$yr$^{-1}$ \citep{kalli:13} -- i.e., 
implying a tangential motion predominantly towards the east. Observations from the Pan-Andromeda Archaeological Survey 
(PAndAS) have revealed the presence of similar stellar streams emanating from the disks of both M33 \citep{mcconnachie:10} 
and NGC 147 \citep[][see also Irwin et al. 2016, in prep.]{crnojevic:14}. These are inferred to have arisen due to recent encounters 
between these satellite galaxies and M31 \citep[e.g.][]{mcconnachie:09}. 

To explore this possibility in more detail we ran several $N$-body simulations of an LMC flyby. These were carried out using 
the $N$-body component of \textsc{gadget-3} which is similar to \textsc{gadget-2} \citep{springel:05}. In each simulation, 
the LMC was modelled as a two component galaxy with a stellar disk and an NFW profile \citep{navarro:97}. We constructed 
self-consistent and stable initial conditions following the procedure described in \citet{villalobos:08} and used the same form 
for the stellar disk as in that work. Before adiabatic contraction, the LMC halo had a mass of $M_{200}=1.2 \times 10^{11} M_\odot$ 
and a concentration of $c_{200} = 8$.  The stellar disk had a mass of $M_d = 4 \times 10^9 M_\odot$, a scale radius of $1.5$\ kpc 
(matching the inferred value in Table 1), and a scale height of $0.3$\ kpc. After the adiabatic contraction of the LMC halo, the 
total LMC mass within $8.7$\ kpc is $1.8 \times 10^{10} M_\odot$, consistent with the mass inferred from the LMC rotation curve 
in \citet{vdm:14}. The total LMC mass is $1.4 \times 10^{11} M_\odot$, consistent with other recent mass estimates 
\citep[e.g.,][]{kalli:13,penarrubia:15}. The initial orientation and rotational sense of the LMC were chosen to match the results 
of \citet{vdm:14}. We used $10^6$ particles each to represent the LMC disk and halo, and a softening length of $75$\ pc and 
$500$\ pc respectively. The Milky Way was modelled as a 3-component system with a bulge, disk, and dark matter halo as 
described in \citet{gomez:15}.

Our $N$-body model was integrated for 2 Gyr. The initial positions and velocities of the Milky Way and LMC were chosen using 
backward integration from the current position as in \citet{gomez:15}, which resulted in the LMC being within 
2$\sigma$ of the present day positions and velocities given in \citet{kalli:13}.

\begin{figure}
\begin{center}
\includegraphics[width=80mm,clip=true]{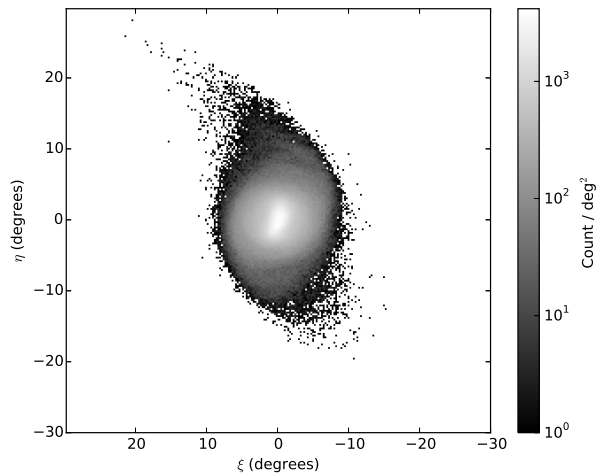}
\end{center}
\caption{Stellar density of our simulated LMC using the same coordinate system as in Figure \ref{f:map}. The northern outskirts of the
disk show a stream of tidally stripped debris qualitatively similar to our observed substructure.
\label{f:sim}}
\end{figure}

Figure \ref{f:sim} shows the density of particles in the disk -- i.e., the stellar density in the simulation -- using the same 
coordinate system as in Figure \ref{f:map}. The simulation shows material being stripped from the northern part of the disk
in the correct direction to give LMC debris near the location of the Carina dwarf -- thus at least qualitatively reproducing the 
properties of our stellar overdensity. In the model the stream debris is due to tidal stripping of the LMC, and the rate of this 
stripping is controlled by both the orientation of the LMC internal rotation and its orbital motion \citep[e.g.,][]{read:06,donghia:09}. 
As a consistency check we re-ran the simulation with the LMC in two orientations, prograde and retograde, and found 
enhanced and decreased stripping respectively, as expected. For the orientation used in our simulation, the direction of the 
LMC spin and the angular momentum of its orbital motion around the Milky Way differ by $75\degr$, so we expect a slight 
enhancement of the stripping over a non-rotating progenitor.  Intriguingly, at the time shown in Figure
\ref{f:sim} our simulated LMC disk is intrinsically elliptical, with an ellipticity $e \approx 0.15$ at radii commensurate with 
those considered by our disk-fitting algorithm in Section \ref{ss:geom}.

We emphasise that these simulations do not {\it prove} that the observed substructure is a debris stream caused by tidal 
interaction between the LMC and the Milky Way, merely that this is a plausible explanation. In the future we plan to run a 
larger suite of models to explore how the debris is affected by varying the properties of the Milky Way and the LMC, and the 
presence of the SMC.

The SMC is certainly a complicating factor, and indeed we cannot rule out that our substructure is due to one or more close 
encounters between the two Magellanic Clouds, rather than simply due to the tidal force of the Milky Way. The SMC sits at
a position angle of $228.9\degr$ relative to the LMC centre, which is broadly in alignment with the direction of the arc of our 
stream-like overdensity. This galaxy is believed to have had multiple close interactions with the LMC in the past -- perhaps 
as recently as a few hundred Myr ago, and perhaps passing as close as a few kpc from the LMC centre. Observations have revealed
a stellar component in the Magellanic Bridge region \citep[e.g.,][]{irwin:85,nidever:13,noel:13,noel:15,skowron:14}, probably
stripped from the SMC by just such an encounter.

\citet{besla:12} model the effects of a very close LMC-SMC interaction, where the orbits are consistent with the observed 
proper motions; amongst their findings is that such an interaction can substantially disturb the LMC disk and even give 
rise to possible stream-like features -- in the particular case they consider this mainly occurs to the north 
and north-east of the LMC centre as we have observed. More recently, \citet{besla:16} have conducted modelling in light of
their discovery of stellar arcs and spiral arms extending to $\sim 8\degr$ north of the LMC, and the absence of comparable
features in the south. They show that repeated close passages of the SMC can lead to precisely this type of asymmetric arrangement.
Together these computations show that, at least in principle, our LMC overdensity could arise due to interaction(s) with the SMC; 
moreover, such encouters might also offer an explanation for the structure we observe in the LMC disk between $9\degr-12\degr$. 
Ultimately, new observations and more detailed modelling will be required to identify the most likely origin for our stream-like 
feature. A key question is whether a matching substructure exists due south of the LMC centre and stretching to the west -- the 
tidal features seen around both M33 and NGC 147 due to the influence of M31 are quite symmetric in this regard, while the
models of \citet{besla:16} show that if the SMC is responsible then the arrangement should be asymmetric (i.e., there should
be no matching arc in the south).

We note that radial velocity observations along our substructure should help determine whether this feature is a stream of 
material stripped from the outer LMC disk, or an overdensity in the disk itself.  If the latter, then it ought to approximately
follow the known disk rotation profile with position angle. If, on the other hand, it is a stream of tidal debris, then our 
$N$-body model predicts a much more constant velocity with position because the stripped material
tends to simply proceed roughly along the LMC orbit rather than continuing to share in the disk rotation pattern.

Finally, we have observed apparent evidence for LMC populations at radii of up to $\approx 18.5$\ kpc ($\approx 20\degr$) to the 
north and north-west. This corroborates the results of previous studies that have spectroscopically detected apparent LMC populations 
in discrete fields at radii of up to $\sim 20\degr$ and over a wide range of azimuthal angles \citep[e.g.,][]{munoz:06,majewski:09};
we are now beginning to create a fully-filled map of these regions.  The nature of these extremely remote stars is not clear
-- while \citet{majewski:09} interpret them as a classical spheroidal halo, our radial density profile to the north-west is well
described by our best-fitting inclined disk model over all radii beyond $\sim 9\degr$ without evidently requiring a separate
halo component. Furthermore, it is not obvious whether these extremely remote stars could still be gravitationally bound to 
the LMC -- we have observed material that may very well be in the process of being tidally stripped from the LMC at a radius
of $\approx 12.5$\ kpc from the galaxy's centre; on the other hand, the kinematic measurements of \citet{vdm:14} suggest 
that the tidal radius of the LMC could be as large as $22.3\pm 5.2$\ kpc, and \citet{belokurov:16} have observed 
possible LMC members extending to even larger radii. Whether or not these outermost populations remain bound to the 
LMC is a particularly interesting question -- if they are not, then the fact that we are still able to observe their presence 
would likely argue for the LMC only having entered the tidal influence of the Milky Way quite recently, such that it is 
possibly still on its first pass around the Galaxy. Further deep imaging and spectrscopy is necessary to explore the properties of
this very diffuse structure.\vspace{-4mm}

\section*{Acknowledgments}
The authors are grateful to Yuri Beletsky for granting permission to reproduce his image of the LMC in Figure \ref{f:map}, and 
to the organisers of the ESO Workshop {\it Satellites and Streams in Santiago} in April 2015, where this project was initially conceived.
We thank the anonymous referee for their thorough and constructive report, which helped significantly improve the paper.
ADM and GDC acknowledge support from the Australian Research Council (Discovery Projects DP120101237 and DP150103294).
The research leading to these results has received funding from the European Research Council under the European Union's Seventh 
Framework Programme (FP/2007-2013)/ERC Grant Agreement no. 308024.  

This project used public archival data from the Dark Energy Survey (DES). 
Funding for the DES Projects has been provided by 
the U.S. Department of Energy, 
the U.S. National Science Foundation, 
the Ministry of Science and Education of Spain, 
the Science and Technology Facilities Council of the United Kingdom, 
the Higher Education Funding Council for England, 
the National Center for Supercomputing Applications at the University of Illinois at Urbana-Champaign, 
the Kavli Institute of Cosmological Physics at the University of Chicago, 
the Center for Cosmology and Astro-Particle Physics at the Ohio State University, 
the Mitchell Institute for Fundamental Physics and Astronomy at Texas A\&M University, 
Financiadora de Estudos e Projetos, Funda{\c c}{\~a}o Carlos Chagas Filho de Amparo {\`a} Pesquisa do Estado do Rio de Janeiro, 
Conselho Nacional de Desenvolvimento Cient{\'i}fico e Tecnol{\'o}gico and the Minist{\'e}rio da Ci{\^e}ncia, Tecnologia e Inovac{\~a}o, 
the Deutsche Forschungsgemeinschaft, 
and the Collaborating Institutions in the Dark Energy Survey. 
The Collaborating Institutions are 
Argonne National Laboratory, 
the University of California at Santa Cruz, 
the University of Cambridge, 
Centro de Investigaciones En{\'e}rgeticas, Medioambientales y Tecnol{\'o}gicas-Madrid, 
the University of Chicago, 
University College London, 
the DES-Brazil Consortium, 
the University of Edinburgh, 
the Eidgen{\"o}ssische Technische Hoch\-schule (ETH) Z{\"u}rich, 
Fermi National Accelerator Laboratory, 
the University of Illinois at Urbana-Champaign, 
the Institut de Ci{\`e}ncies de l'Espai (IEEC/CSIC), 
the Institut de F{\'i}sica d'Altes Energies, 
Lawrence Berkeley National Laboratory, 
the Ludwig-Maximilians Universit{\"a}t M{\"u}nchen and the associated Excellence Cluster Universe, 
the University of Michigan, 
{the} National Optical Astronomy Observatory, 
the University of Nottingham, 
the Ohio State University, 
the University of Pennsylvania, 
the University of Portsmouth, 
SLAC National Accelerator Laboratory, 
Stanford University, 
the University of Sussex, 
and Texas A\&M University. \vspace{-3mm}












\bsp	
\label{lastpage}
\end{document}